\documentclass[12pt]{article}
\def\slash#1{\setbox0=\hbox{$#1$}#1\hskip-\wd0\hbox to\wd0{\hss\sl/\/\hss}}
\usepackage{epsf, cite, amsmath, amssymb}
\usepackage{epsfig}
\usepackage[all]{xy}
\makeatletter
\renewcommand\section{\@startsection {section}{1}{\z@}%
                                   {-3.5ex \@plus -1ex \@minus -.2ex}
                                   {2.3ex \@plus.2ex}%
                                   {\normalfont\large\bfseries}}
\renewcommand\subsection{\@startsection{subsection}{2}{\z@}%
                                     {-3.25ex\@plus -1ex \@minus -.2ex}%
                                     {1.5ex \@plus .2ex}%
                                     {\normalfont\bfseries}}
\makeatother

\let\non\nonumber

\newcommand{\bea}{\begin{eqnarray}}
\newcommand{\eea}{\end{eqnarray}}
\newcommand{\be}{\begin{equation}}
\newcommand{\ee}{\end{equation}}

\newcommand{\p}{\partial}

\newcommand{\s}{\sigma}

\newcommand{\C}[1]{$(\ref{#1})$}

\begin{document}
\begin{titlepage}

\begin{center}



\vskip 2 cm
{\Large \bf The $D^6 {\cal{R}}^4$ term in type IIB string theory on $T^2$ and U--duality}\\
\vskip 1.25 cm { Anirban Basu\footnote{email: abasu@ias.edu}
}\\
{\vskip 0.75cm
Institute for Advanced Study, Princeton, NJ 08540, USA\\
}

\end{center}

\vskip 2 cm

\begin{abstract}
\baselineskip=18pt

We propose a manifestly U--duality invariant modular form for the $D^6 {\cal{R}}^4$ interaction 
in the effective action of type IIB string theory compactified on $T^2$. It receives perturbative
contributions upto genus three, as well as non--perturbative contributions from D--instantons and 
$(p,q)$ string instantons wrapping $T^2$. Our construction is based on constraints coming from string 
perturbation theory, U--duality, the decompactification limit to ten 
dimensions, and the equality of the perturbative part of the amplitude in type IIA and type IIB 
string theories. Using duality, parts of the perturbative amplitude are also shown to match exactly the results 
obtained from eleven dimensional supergravity compactified on $T^3$ at one loop. We also obtain parts of the
genus one and genus $k$ amplitudes for the $D^{2k} {\cal{R}}^4$ interaction for arbitrary $k \geq 4$. We enhance a
part of this amplitude to a U--duality invariant modular form.

\end{abstract}

\end{titlepage}

\pagestyle{plain}
\baselineskip=18pt

\section{Introduction}

It is an important problem to construct the low energy effective action of string theory. Not only
does it yield valuable information about the perturbative and non--perturbative structure of string theory,
but is also elucidates the role of U--duality. The effective action of string theory can be constructed
perturbatively in $\alpha'$, the inverse of the string tension. Of course there are also expected to be 
corrections which are non--perturbative in $\alpha'$. Constructing certain interactions in the
effective action is sometimes tractable in theories with maximal supersymmetry. These special interactions
are BPS, and receive only a finite number of perturbative contributions, as well as corrections due
to various instantons. We shall consider the special case of toroidal compactification of type IIB superstring 
theory to eight dimensions, such that it preserves all the thirty two
supersymmetries. 

Certain classes of BPS interactions in the low energy eight dimensional effective action
are expected to satisfy non--renormalization theorems. For example, the $D^{2k} {\mathcal{R}}^4$ 
interactions (at least for sufficiently low values of $k$), where $k$ is a 
non--negative integer, are expected to receive only a finite number of perturbative contributions, as well as
non--perturbative corrections from D--instantons, and $(p,q)$ string instantons wrapping $T^2$. 
Here ${\cal{R}}^4$ stands for the $t_8 t_8 R^4$ interaction~\cite{Green:1981xx,Green:1981yb,D'Hoker:1988ta}, 
and can be expressed entirely in terms of four powers of the Weyl tensor.   
The U--duality symmetry and maximal supersymmetry imposes strong constraints on these interactions.

Type IIB superstring theory compactified on $T^2$ has a conjectured U--duality symmetry group
$SL(2,\mathbb{Z})_U \times SL(3,\mathbb{Z})_M$~\cite{Hull:1994ys,Witten:1995ex}. The complex structure
modulus $U$ of $T^2$ transforms non--trivially under $SL(2,\mathbb{Z})_U$ as
\be U \rightarrow \frac{a U + b}{c U + d}, \ee
where $a,b,c,d \in \mathbb{Z}$, and $ad -bc =1$.

The $SL(3,\mathbb{Z})_M$
factor of the U--duality group arises in a somewhat involved way. The theory has an 
$SL(2,\mathbb{Z})_\tau$ (S--duality) symmetry 
under which the complexified coupling 
\be \tau = \tau_1 + i \tau_2 = C_0 + i e^{-\phi}, \ee
transforms as
\be \tau \rightarrow \frac{a \tau + b}{c \tau + d} ,\ee
while the combination $B_R + \tau B_N$ transforms as
\be B_R + \tau B_N \rightarrow \frac{B_R + \tau B_N}{c\tau + d}, \ee
where $B_N (B_R)$ is the modulus from the NS--NS (R--R) two form on $T^2$.
It also has an $SL(2,\mathbb{Z})_T$ (T--duality) symmetry under which 
the Kahler structure modulus of $T^2$  
\be T = B_N + i V_2 ,\ee
transforms as
\be T \rightarrow \frac{a T + b}{c T + d}, \ee
where $V_2$ is the volume of $T^2$ in the 
string frame. It also acts on the complex scalar $\rho$ defined by
\be  \rho = - B_R + i \tau_1 V_2 , \ee
as \be \rho \rightarrow \frac{\rho}{c T + d}, \ee
while leaving the eight dimensional dilaton invariant. The $SL(2,\mathbb{Z})_\tau$ and $SL(2,\mathbb{Z})_T$ symmetries 
can be intertwined and embedded into the $SL(3,\mathbb{Z})_M$ factor of the U--duality group.
 
The part of the supergravity action involving the scalars can be written in the Einstein frame as 
(we are following the conventions of~\cite{Kiritsis:1997em}) 
\be \label{act8}
S \sim \frac{1}{l_s^6} \int d^{8} x \sqrt{-\hat{g}_8} \Big( - \frac{\p_\mu U \hat{\p}^\mu 
\bar{U}}{2 U_2^2} + \frac{1}{4} {\rm Tr} (\p_\mu M \hat{\p}^\mu M^{-1}) + \ldots \Big),\ee 
where the hat denotes quantities in the eight dimensional 
Einstein frame. In \C{act8}, $M$ is a symmetric matrix with determinant one given by
\be \label{matrix}
M = \nu^{1/3} \begin{pmatrix} 1/\tau_2 & \tau_1/\tau_2 & {\rm Re}(B)/\tau_2 \\ \tau_1/\tau_2& \vert 
\tau \vert^2/\tau_2 & {\rm Re}(\bar{\tau} B)/\tau_2  \\
{\rm Re}(B)/\tau_2 & {\rm Re}(\bar{\tau} B)/\tau_2 & 1/\nu + \vert B \vert^2/\tau_2  \end{pmatrix}, \ee
where $B = B_R + \tau B_N$, and $\nu = (\tau_2 V_2^2)^{-1}$.

In the Einstein
frame, where the metric is U--duality invariant, the coefficients of these protected
$D^{2k} {\mathcal{R}}^4$ interactions
should be given by modular forms of the U--duality group, which are invariant under  
$SL(2,\mathbb{Z})_U \times SL(3,\mathbb{Z})_M$ transformations. Constructing these modular forms 
for toroidal compactifications of type II string theory and M theory that
preserve maximal supersymmetry, and analyzing their
non--renormalization properties have been worked out for some of these operators in various 
dimensions~\cite{Green:1997tv,Green:1997di,Green:1997as,Kiritsis:1997em,Pioline:1997pu,
Pioline:1998mn,Green:1998by,Obers:1998rn,Obers:1999um,Green:1999pu,Obers:1999es,
Pioline:2001jn,Berkovits:2004px,D'Hoker:2005jc,D'Hoker:2005ht,Green:2005ba,Berkovits:2006vc,Green:2006gt} 
(see~\cite{Obers:1998fb,Green:1999qt} for reviews). In eight dimensions, a modular form for the 
$D^4 {\mathcal{R}}^4$ interaction has been proposed recently~\cite{Basu:2007ru}. In this work, we 
shall propose a manifestly U--duality invariant modular form for the $D^6 {\mathcal{R}}^4$ interaction
in the effective action. By this, we actually mean the
\be (s^3 + t^3 + u^3) {\cal{R}}^4 \ee 
interaction involving the elastic scattering of two gravitons.

To summarize, we propose that modular form is given by
\bea \label{finexp}
&&{\mathcal{E}}_{(3/2,3/2)} (M) +\frac{20}{3}
E_{3} (M^{-1})^{SL(3,\mathbb{Z})} E_{3} (U, \bar{U})^{SL(2,\mathbb{Z})} \non \\ &&
+ f (U, \bar{U}) +\frac{1}{2}
E_{3/2} (M)^{SL(3,\mathbb{Z})} E_{1} (U, \bar{U})^{SL(2,\mathbb{Z})} ,
\eea
where $E_{s} (M)^{SL(3,\mathbb{Z})}$ ($E_{s} (M^{-1})^{SL(3,\mathbb{Z})}$) is the non--holomorphic modular invariant
Eisenstein series of ${SL(3,\mathbb{Z})}_M$ of order $s$ in the fundamental 
(anti--fundamental) representation.
Also $E_{s} (U, \bar{U})^{SL(2,\mathbb{Z})}$ is the non--holomorphic modular invariant
Eisenstein series of ${SL(2,\mathbb{Z})}_U$. These Einstein series satisfy the Laplace equation on the 
fundamental domain of moduli space. On the other hand, $f (U, \bar{U})$ and ${\mathcal{E}}_{(3/2,3/2)} (M)$
are ${SL(2,\mathbb{Z})}_U$ and ${SL(3,\mathbb{Z})}_M$ invariant modular forms respectively,
that satisfy Poisson equation on the fundamental domain of moduli space given by
\be \Delta_{{SL(2,\mathbb{Z})}_U} f (U, \bar{U}) = 12 f (U, \bar{U}) -6
\Big( E_1 (U, \bar{U}) \Big)^2, \ee
and
\be \Delta_{SL(3,\mathbb{Z})} {\mathcal{E}}_{(3/2,3/2)} (M) = 12 {\mathcal{E}}_{(3/2,3/2)} (M)
-\frac{3}{2} \Big( E_{3/2} (M) \Big)^2.\ee

We begin by constructing the perturbative part of the modular form. Constraints coming from
string perturbation theory, U--duality, the decompactification limit to
ten dimensions, and the equality of the 
perturbative part of the amplitude in type IIA and type IIB string theories, lead us to 
propose the complete perturbative part of the modular form.\footnote{Since the ${\mathcal{R}}^4$ interaction
involves the even--even spin structures only, the perturbative contributions have to be 
the same in the two type II string theories. Thus this part of the amplitude must be symmetric
under the interchange of $U$ and $T$, while the eight dimensional IIA dilaton goes to the IIB dilaton and 
vice versa.} This receives contributions only
upto genus three in string perturbation theory. Using duality, we next provide evidence 
for some of these contributions by analyzing the one loop
four graviton scattering amplitude in eleven dimensional supergravity compactified on 
$T^3$. 

We next propose the exact expression for the modular form based on constraints of supersymmetry
and the ten dimensional $SL(2,\mathbb{Z})_\tau$ invariant answer. 
This provides the non--perturbative completion of the perturbative
part of the modular form, and involves contributions from D--instantons, as well as from 
$(p,q)$ string instantons wrapping $T^2$.  Analyzing one loop eleven dimensional supergravity 
compactified on $T^3$, we also obtain parts of the
genus one and genus $k$ amplitudes for the $D^{2k} {\cal{R}}^4$ interaction for arbitrary $k \geq 4$. 
We enhance a part of this amplitude to a U--duality invariant modular form.
We also make some comments about generalizing our construction to toroidal compactifications
with maximal supersymmetry to lower dimensions. 
In the appendices, relevant details for the Eisenstein series of $SL(2,\mathbb{Z})$ and 
$SL(3,\mathbb{Z})$, and the torus amplitude are summarised. They also contain a discussion about
possible contributions to the modular form we might have missed, where we provide arguments 
that they should vanish.

\section{The perturbative part of the proposed modular form}

We begin by constructing the perturbative part of the proposed modular form.
The low energy effective action for type IIB superstring theory in ten dimensions includes the interaction
(in the string frame)~\cite{Green:2005ba} 
\be \label{term10d}
S \sim l_s^4 \int d^{10} x \sqrt{-g}  \Big(\zeta(3)^2 e^{-2\phi} + 2 \zeta (3) \zeta (2) + 
6 \zeta (4) e^{2\phi} + \frac{2}{9} \zeta (6) e^{4\phi} + \ldots \Big) D^6 {\cal{R}}^4, \ee
where the $\ldots$ involve contributions from D--instantons. Thus from \C{term10d}, we see that the
$D^6 {\cal{R}}^4$ interaction receives perturbative contributions only upto genus three. 
Compactifying on $T^2$ of volume 
$V_2 l_s^2$ in the string frame,
this leads to an interaction in the eight dimensional Einstein frame given by
\be S \sim l_s^6 \int d^{8} x \sqrt{-\hat{g}_8}  \Big( V_2 e^{-\phi} \Big)^2 \Big(
\zeta(3)^2 e^{-2\phi} + 2 \zeta (3) \zeta (2) + 
6 \zeta (4) e^{2\phi} + \frac{2}{9} \zeta (6) e^{4\phi} + \ldots
\Big) \hat{D}^6 \hat{\cal{R}}^4 .\ee
Thus the modular form for the $D^6 {\mathcal{R}}^4$ interaction
must include, among other terms,
\be \Big( V_2 e^{-\phi} \Big)^2 \Big(  \zeta(3)^2 e^{-2\phi} + 2 \zeta (3) \zeta (2) + 
6 \zeta (4) e^{2\phi} + \frac{2}{9} \zeta (6) e^{4\phi} \Big). \ee
We first construct the perturbative part of the modular form.

\subsection{Constraints using string perturbation theory}

Let us consider the perturbative contributions to the $D^6 {\cal{R}}^4$ interaction.
As mentioned before, by this interaction, we actually mean the term
\be (s^3 + t^3 + u^3) {\cal{R}}^4 \ee 
in the four graviton scattering amplitude.

Consider the tree level and one loop
amplitudes for this interaction using string perturbation theory.
The sum of the contributions to the four graviton amplitude at tree level~\cite{Green:1981xx,D'Hoker:1988ta} 
and at one loop~\cite{Green:1981ya,D'Hoker:1988ta} in type II 
string theory compactified on $T^2$ is proportional to\footnote{The calculation actually
yields ${\cal{R}}^4$ at the linearized level.} 
\be \label{totalcont}
\Big[- V_2 e^{-2 \phi} \frac{\Gamma (-l_s^2 s/4) \Gamma (-l_s^2 t/4) 
\Gamma (-l_s^2 u/4)}{ \Gamma (1 + l_s^2 s/4) \Gamma (1 + l_s^2 t/4)
\Gamma (1 + l_s^2 u/4)} + 2\pi I \Big] {\cal{R}}^4,\ee
where $V_2$ is the volume of $T^2$ in the string frame, $s,t,u$ are the Mandelstam variables
, and $I$ is obtained from the one loop amplitude. We are looking at the part of the amplitude
involving the even--even spin structures, and hence the amplitude is the same for type IIA and type IIB
string theories. Now $I$ is given by
\be \label{defd1}
I = \int_{\cal{F}} \frac{d^2 \Omega}{\Omega_2^2} Z_{lat} F(\Omega,\bar\Omega) ,\ee
where $\cal{F}$ is the fundamental domain of $SL(2,\mathbb{Z})$, and $d^2 \Omega = d\Omega d \bar\Omega/2$. 
The relative coefficient between the tree level and the one loop
terms in \C{totalcont} is fixed using unitarity~\cite{Sakai:1986bi}. 
In \C{defd1}, the lattice factor $Z_{lat}$ which depends on the moduli is given by~\cite{Dixon:1990pc}
\bea \label{deflattice}
Z_{lat} &=& V_2 \sum_{m_1,m_2,n_1,n_2 \in \mathbb{Z}} e^{-\frac{\pi}{\Omega_2} \sum_{i,j} (G + B_N)_{ij}
(m_i + n_i \Omega) (m_j + n_j \bar\Omega)} \non \\ &=& V_2 \sum_{A \in {Mat} (2 \times 2, \mathbb{Z})}
{\rm exp} \Big[ -2\pi i T ({\rm det} A) - \frac{\pi T_2}{\Omega_2 U_2} 
\Big\vert \begin{pmatrix} 1 & U \end{pmatrix}
A \begin{pmatrix} \Omega \\ 1 
\end{pmatrix} \Big\vert^2 \Big], \eea
where 
\be G_{ij} = \frac{T_2}{U_2} \begin{pmatrix} 1 & U_1 \\ U_1 & \vert U \vert^2 
\end{pmatrix}. \ee

Also the dynamical factor $F(\Omega,\bar\Omega)$ in \C{defd1}, which is independent of the
moduli, is given by
\be  \label{factF}
F(\Omega,\bar\Omega) = \int_{\cal{T}} \prod_{i=1}^3 \frac{d^2 \nu_i}{\Omega_2} 
(\chi_{12} \chi_{34})^{l_s^2 s} (\chi_{14} \chi_{23})^{l_s^2 t} (\chi_{13} \chi_{24})^{l_s^2 u}. \ee

In \C{factF}, $\nu_i$ ($i=1,\ldots,4$) are the positions of insertions of
the four vertex operators on the toroidal 
worldsheet, 
and $\nu_4$ has been set equal to $\Omega$ using conformal invariance. Also $d^2 \nu_i = d \nu_i^R d\nu_i^I$,
where $\nu_i^R$ ($\nu_i^I$) are the real (imaginary) parts of $\nu_i$. The integral over $\cal{T}$ is over
the domain ${\cal{T}} = \{ -1/2 \leq \nu_i^R < 1/2 ,  0 \leq \nu_i^I < \Omega_2 \}$.
Finally, ${\rm ln } \chi (\nu_i - \nu_j ; \Omega)$ is the scalar Green function between the points
$\nu_i$ and $\nu_j$ on the toroidal worldsheet.

Expanding \C{defd1} to sixth order in the momenta, we get that
\be I = \frac{l_s^6}{3} (s^3 + t^3 + u^3) [\hat{I}_1 + \hat{I}_2] ,\ee

where
\be \label{torone}
\hat{I}_1 = 4 \int_{{\cal{F}}_L} \frac{d^2 \Omega}{\Omega_2^2} Z_{lat} \int_{\cal{T}} \prod_{i=1}^3 \frac{d^2 
\nu_i}{\Omega_2}  {\rm ln } \hat\chi (\nu_1 - \nu_2 ; 
\Omega) {\rm ln } \hat\chi (\nu_1 - \nu_3 ;\Omega) \hat\chi (\nu_2 - \nu_3 ;\Omega)
, \ee

and
\be \label{tortwo}
\hat{I}_2 = \int_{{\cal{F}}_L} \frac{d^2 \Omega}{\Omega_2^2} Z_{lat} \int_{\cal{T}} \prod_{i=1}^3 \frac{d^2 
\nu_i}{\Omega_2}  [ {\rm ln } \hat\chi (\nu_1 - \nu_2 ; 
\Omega)]^3, \ee

which can be depicted diagrammatically as in Figure 1.

\begin{figure}[ht]
\[
\mbox{\begin{picture}(195,85)(10,10)
\includegraphics[scale=0.40]{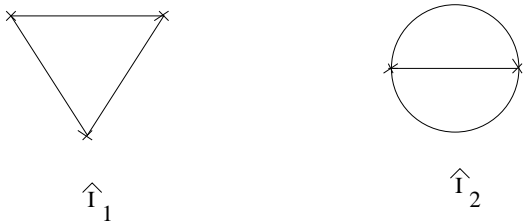}
\end{picture}}
\]
\caption{\it
Schematics of the torus amplitude.
}
\end{figure}

In the expressions above, we have defined
\be \hat\chi (\nu_i - \nu_i ;\Omega) = \chi (\nu_i - \nu_j ;\Omega) - \frac{1}{2} {\rm ln} \Big\vert 
(2\pi)^{1/2}\eta(\Omega) \Big\vert^2 .\ee

Thus we have removed the zero mode part of the scalar propagator, which does not contribute to the
on--shell amplitude using $s+t+u =0$. 

In \C{torone} and \C{tortwo}, 
note that the one loop contribution has been integrated 
over the restricted fundamental domain 
${\cal{F}}_L$ of $SL(2,\mathbb{Z})$, which is obtained from $\cal{F}$ by restricting to $\Omega_2 
\leq L$. This is necessary to separate the analytic parts of the amplitude from the non--analytic parts
(see~\cite{Green:1999pv} for a detailed discussion).
The integral over ${\cal{F}}_L$ gives both finite and divergent terms to the amplitude in the 
limit $L \rightarrow \infty$. The terms which are finite in this limit are the 
analytic parts of the amplitude. 
The parts which diverge in this limit cancel in the whole amplitude when the contribution from the part of 
the moduli space $\cal{F}$ with $\Omega_2 > L$ is also included. In addition to 
these divergences which cancel,
the contribution from $\cal{F}$ with $\Omega_2 > L$ also gives the various 
non--analytic terms in the amplitude.
Keeping this in mind, we shall consider only the contributions which are finite in the limit 
$L \rightarrow \infty$ and drop all divergent terms. In the calculations,
we shall see that the domain of integration ${\cal{F}}$ shall often be changed to the upper half plane or a 
strip. Then truncating to ${\cal{F}}_L$ to calculate the analytic terms cannot be done when the
integration over ${\cal{F}}_L$ produces divergences of the form ${\rm ln} L$~\cite{Green:1999pv}. However,
for our case there are no logarithmic divergences, and so this is not a problem for us.

In calculating both $\hat{I}_1$ and $\hat{I}_2$, we need to add the contributions from the zero orbit, 
the non--degenerate orbits and the degenerate orbits of $SL(2,\mathbb{Z})$ respectively~\cite{Dixon:1990pc}.

{\bf{(i)}} The contribution from the zero orbit involves setting $A =0$ in \C{deflattice}.

{\bf{(ii)}} The contribution from the non--degenerate orbits involves setting
\be 
A =  \begin{pmatrix} k & j  \\ 0 & p 
\end{pmatrix} ,\ee
where $ k > j \geq 0, p \neq 0$ in \C{deflattice}, and changing the domain of integration to be the 
double cover of the upper half plane.

{\bf{(iii)}} The contribution from the degenerate orbits involves setting
\be 
A =  \begin{pmatrix} 0 & j  \\ 0 & p 
\end{pmatrix} \ee
such that $(j,p) \neq (0,0)$ in \C{deflattice}, and changing the domain of integration to be the strip 
$0 < \Omega_2 < L, \vert \Omega_1 \vert < 1/2.$

The details of the calculation of $\hat{I}_1$ and $\hat{I}_2$ are given in the appendix. This
gives us
\bea \hat{I}_1 &=& \frac{1}{8\pi^6} E_{3} (U, \bar{U})^{SL(2,\mathbb{Z})}
E_{3} (T, \bar{T})^{SL(2,\mathbb{Z})}, \non \\ \hat{I}_2 &=& 
\frac{1}{32 \pi^6} E_{3} (U, \bar{U})^{SL(2,\mathbb{Z})}
E_{3} (T, \bar{T})^{SL(2,\mathbb{Z})} \non \\ &&+ \frac{3}{32\pi^3} \zeta (2) \zeta (3) \Big( 
E_{1} (U, \bar{U})^{SL(2,\mathbb{Z})} + E_{1} (T, \bar{T})^{SL(2,\mathbb{Z})}\Big).\eea

Thus the total amplitude in \C{totalcont} gives
\bea \label{partpert}
&&\Big[ \zeta(3)^2 e^{-2\phi} V_2 + \frac{10}{\pi^5} E_{3} (U, \bar{U})^{SL(2,\mathbb{Z})}
E_{3} (T, \bar{T})^{SL(2,\mathbb{Z})} \non \\ &&+ \zeta(3) \Big( 
E_{1} (U, \bar{U})^{SL(2,\mathbb{Z})} + E_{1} (T, \bar{T})^{SL(2,\mathbb{Z})}\Big)
\Big] l_s^6 (s^3 + t^3 + u^3) {\mathcal{R}}^4. \eea

\subsection{Constraints using U--duality and the decompactification limit}

Having obtained the tree level and the one loop contributions to the scattering amplitude, we now show 
how U--duality and the decompactification limit constrains the perturbative structure of the modular 
form. Now \C{partpert}  leads to the term in the effective action in the Einstein frame given by
\bea l_s^6 \int d^{8} x \sqrt{-\hat{g}_8} V_2 e^{-2\phi}  \Big[
\zeta(3)^2 e^{-2\phi} V_2 + \frac{10}{\pi^5} E_{3} (U, \bar{U})^{SL(2,\mathbb{Z})}
E_{3} (T, \bar{T})^{SL(2,\mathbb{Z})} \non \\ + \zeta(3) \Big( 
E_{1} (U, \bar{U})^{SL(2,\mathbb{Z})} + E_{1} (T, \bar{T})^{SL(2,\mathbb{Z})}\Big)
\Big] \hat{D}^6 \hat{\cal{R}}^4 .\eea

Thus the tree level and the one loop contributions to the modular form are given by
\bea \label{impUd}
&&\zeta(3)^2 \Big( \tau_2^2 V_2 \Big)^2 + \tau_2^2 V_2\Big[ \frac{10}{\pi^5} 
E_{3} (U, \bar{U})^{SL(2,\mathbb{Z})}
E_{3} (T, \bar{T})^{SL(2,\mathbb{Z})} \non \\ &&+ \zeta(3) \Big( 
E_{1} (U, \bar{U})^{SL(2,\mathbb{Z})} + E_{1} (T, \bar{T})^{SL(2,\mathbb{Z})}\Big) \Big]. \eea 

Note that the $U$ dependent parts of the modular form in \C{impUd} involving 
$E_{3} (U, \bar{U})^{SL(2,\mathbb{Z})}$ and $E_{1} (U, \bar{U})^{SL(2,\mathbb{Z})}$ are 
$SL(2,\mathbb{Z})_U$ invariant. Thus whatever multiplies these terms must be $SL(3,\mathbb{Z})_M$
invariant. Thus in \C{impUd}, the two expressions
\be \label{enhan1}
\tau_2^2 V_2 E_{3} (T, \bar{T})^{SL(2,\mathbb{Z})} \ee
which multiplies $E_{3} (U, \bar{U})^{SL(2,\mathbb{Z})}$, and 
\be \label{enhan2}
\tau_2^2 V_2 \ee 
which multiplies $E_{1} (U, \bar{U})^{SL(2,\mathbb{Z})}$, must both be enhanced to
invariant modular forms of $SL(3,\mathbb{Z})_M$. Such modular forms need not be simple expressions
involving Eisenstein series of $SL(3,\mathbb{Z})_M$. For example, the modular forms for the 
${\mathcal{R}}^4$ and the $D^4 {\mathcal{R}}^4$ interactions in ten dimensions are given by Eisenstein series
of $SL(2,\mathbb{Z})_\tau$ which satisfies the Laplace equation on the fundamental domain of
$SL(2,\mathbb{Z})_\tau$, however, the modular form for the $D^6 {\mathcal{R}}^4$ interaction 
is more complicated, and satisfies a Poisson equation on the fundamental domain of
$SL(2,\mathbb{Z})_\tau$. However, we now argue that there are simple and natural modular forms
of $SL(3,\mathbb{Z})_M$ to which \C{enhan1} and \C{enhan2} can be ehnanced to.
 
In order to motivate natural candidates for these modular forms, from \C{impUd} note that
the genus $g$ contribution to the perturbative part of the modular form involves $(\tau_2^2 
V_2 )^{2-g}$. Given the structure of the perturbative contributions to $E_s (M)^{SL(3,\mathbb{Z})}$
which follow from \C{expEs}, we see that the possible choices are severely restricted. In fact, there 
are only two possibilities: 

(i) $E_{-3/2} (M)^{SL(3,\mathbb{Z})}$ which contributes at genus one and three, and

(ii) $E_{3/2} (M)^{SL(3,\mathbb{Z})}$ which contributes at genus one and two.

The only other possibility based on the $\tau_2^2 V_2$ dependence is $E_{-9/2} (M)^{SL(3,\mathbb{Z})}$
which contributes at genus zero and five. However the tree level contribution is proportional to
$(\tau_2^2 V_2 )^2 E_{6} (T, \bar{T})^{SL(2,\mathbb{Z})}$, which is inconsistent with the known tree level 
amplitude.  

In fact, from \C{expEs}, we see that\footnote{We use $\zeta (-3) =1/120$.}
\be E_{-3/2} (M)^{SL(3,\mathbb{Z})}_{\rm pert} = \frac{1}{60} \Big( \tau_2^2 V_2 \Big)^{-1}
 + \frac{3}{2\pi^5}  \tau_2^2 V_2  E_{3} (T, \bar{T})^{SL(2,\mathbb{Z})} ,\ee
which has a genus one contribution involving \C{enhan1}, where we have also used 
the relation \C{relSL2}. Also we have that
\be E_{3/2} (M)^{SL(3,\mathbb{Z})}_{\rm pert} = 2\zeta (3) \tau_2^2 V_2 
 + 2E_{1} (T, \bar{T})^{SL(2,\mathbb{Z})} ,\ee
which has a genus one contribution involving \C{enhan2}. This suggests a natural enhancement
\bea \frac{10}{\pi^5} \tau_2^2 V_2 E_{3} 
(T, \bar{T})^{SL(2,\mathbb{Z})} E_{3} (U, \bar{U})^{SL(2,\mathbb{Z})} \rightarrow \frac{20}{3}
E_{-3/2} (M)^{SL(3,\mathbb{Z})}_{\rm pert} E_{3} (U, \bar{U})^{SL(2,\mathbb{Z})} , \non \\
\zeta (3) \tau_2^2 V_2 E_{1} (U, \bar{U})^{SL(2,\mathbb{Z})} \rightarrow \frac{1}{2}
E_{3/2} (M)^{SL(3,\mathbb{Z})}_{\rm pert} E_{1} (U, \bar{U})^{SL(2,\mathbb{Z})}. \eea

Thus \C{impUd} gets enhanced to 
\bea \label{almcomp}
\zeta(3)^2 \Big( \tau_2^2 V_2 \Big)^2 + \zeta(3) \tau_2^2 V_2 
E_{1} (T, \bar{T})^{SL(2,\mathbb{Z})} + \frac{1}{9} \Big( \tau_2^2 V_2 \Big)^{-1}
E_{3} (T, \bar{T})^{SL(2,\mathbb{Z})} 
\non \\ +\frac{20}{3}
E_{-3/2} (M)^{SL(3,\mathbb{Z})}_{\rm pert} E_{3} (U, \bar{U})^{SL(2,\mathbb{Z})}
+\frac{1}{2}
E_{3/2} (M)^{SL(3,\mathbb{Z})}_{\rm pert} E_{1} (U, \bar{U})^{SL(2,\mathbb{Z})} ,
\eea
where we have added the term involving $( \tau_2^2 V_2 )^{-1} E_{3} (T, \bar{T})^{SL(2,\mathbb{Z})}$
by hand. This is a genus three contribution and has to be added to ensure the perturbative equality 
of the type IIA and type IIB scattering amplitudes, for reasons explained before. 

However as we shall explain below, \C{almcomp} cannot be the complete perturbative part of the modular 
form, because it does not give the correct perturbative contributions on decompactifying to ten dimensions:
the genus two contribution vanishes as we shall shortly explain, contradicting \C{term10d}. We thus add a term 
\be \label{addcomp} f(T, \bar{T}) + f(U,\bar{U})\ee
by hand to \C{almcomp}, where $f(T,\bar{T})$ ($f(U,\bar{U})$) is invariant under $SL(2,\mathbb{Z})_T$ 
($SL(2,\mathbb{Z})_U$) transformations. 
This yields a genus two contribution, and is also manifestly symmetric under interchange of $T$
and $U$. We shall fix $f(T,\bar{T})$ later.  

Thus adding \C{almcomp} and \C{addcomp}, we propose that the complete perturbative part of the
modular form is given by
\bea \label{totcomp}
&&\zeta(3)^2 \Big( \tau_2^2 V_2 \Big)^2 + \zeta(3) \tau_2^2 V_2 
E_{1} (T, \bar{T})^{SL(2,\mathbb{Z})}  + f(T,\bar{T}) + \frac{1}{9} \Big( \tau_2^2 V_2 \Big)^{-1}
E_{3} (T, \bar{T})^{SL(2,\mathbb{Z})}
\non \\ &&+ f(U,\bar{U}) +\frac{20}{3}
E_{-3/2} (M)^{SL(3,\mathbb{Z})}_{\rm pert} E_{3} (U, \bar{U})^{SL(2,\mathbb{Z})}
+\frac{1}{2}
E_{3/2} (M)^{SL(3,\mathbb{Z})}_{\rm pert} E_{1} (U, \bar{U})^{SL(2,\mathbb{Z})} .
\eea

Thus, converting to the string frame, we see that 
\C{totcomp} yields the contributions
\bea \label{genusval}
&&{\rm genus}~0 : \zeta(3)^2 ,\non \\ &&{\rm genus} ~1 : 
\frac{10}{\pi^5} 
E_{3} (U, \bar{U})^{SL(2,\mathbb{Z})} E_{3} (T, \bar{T})^{SL(2,\mathbb{Z})} 
+ \zeta(3) \Big( E_{1} (U, \bar{U})^{SL(2,\mathbb{Z})} 
+ E_{1} (T, \bar{T})^{SL(2,\mathbb{Z})}
\Big) ,\non \\ && {\rm genus} ~2 : E_{1} (U, \bar{U})^{SL(2,\mathbb{Z})} E_{1} 
(T, \bar{T})^{SL(2,\mathbb{Z})} + f(T, \bar{T}) + f(U,\bar{U}),\non \\ && {\rm genus} ~3 : \frac{1}{9}
\Big( E_{3} (U, \bar{U})^{SL(2,\mathbb{Z})} 
+ E_{3} (T, \bar{T})^{SL(2,\mathbb{Z})} \Big) 
,\eea
and so the perturbative part of the amplitude is the same in type IIA and type IIB string theories. 
   
We now show that in ten dimensions, \C{totcomp} without the $f(T, \bar{T}) + f(U,\bar{U})$ term, gives all the
contributions in \C{term10d} except the genus two contribution. 
We first decompactify to nine dimensions by defining 
\be T_2 = r_\infty r_B , \quad U_2 = \frac{r_\infty}{r_B} ,\ee 
where $r_\infty$ is the direction that is being decompactified. Here $r_\infty$ and $r_B$ are the radii
of $T^2$ in the string frame. Now let us take the limit $r_\infty \rightarrow \infty$, so that
$T_2, U_2 \rightarrow \infty$. This leads to the nine dimensional interaction
\bea \label{decomp9}
l_s^5 \int d^9 x \sqrt{-g_9} \Big[ (r_B e^{-2 \phi}) \zeta (3)^2 + \Big\{
\frac{15}{\pi^4} \zeta (5) \zeta (6)
\Big( r_B^5 + \frac{1}{r_B^5}\Big) + 2 \zeta (2) \zeta (3) \Big( r_B + \frac{1}{r_B} \Big)
\Big\} \non \\ + 4 \zeta (2)^2 (r_B e^{-2 \phi})^{-1}+ \frac{2}{9} \zeta (6) (r_B e^{-2 \phi})^{-2} 
\Big( r_B^3 + \frac{1}{r_B^3} \Big)\Big] D^6 {\mathcal{R}}^4 ,\eea
where we have set $ l_s \int d^8 x \sqrt{- g_8} r_\infty = \int d^9 x \sqrt{-g_9}$. We have dropped a
term that diverges in the nine dimensional limit. This term comes from the genus one amplitude 
and is given by
\be \label{divnine}
\frac{40}{\pi^5} \zeta (6)^2 l_s^5 \int d^9 x \sqrt{-g_9} r_\infty^5 D^6 {\mathcal{R}}^4 .\ee
This term is only one of an infinite number of such diverging terms coming from the infinite
number of analytic terms. These diverging terms as well as the non--analytic terms must 
add up to give the massless threshold singularity in nine dimensions, and hence do not form a part of the
$D^6 {\mathcal{R}}^4$ interaction in nine dimensions. Clearly because the infinite number of 
divergent terms must add to give the threshold singularity, every divergent term must be independent
of the dilaton, and hence must come from the decompactification limit of the genus one amplitude only.
The fact that there are no divergent terms from the higher genus amplitudes is a consistency check
of our proposal. Also, note that the one loop amplitude in \C{decomp9} precisely
agrees with string perturbation theory~\cite{GRV}, providing a non--trivial check for our proposed
modular form.

Finally, taking the limit $r_B \rightarrow \infty$, we get the term in the ten dimensional effective 
action
\be \label{decomp10} l_s^4 \int d^{10} x \sqrt{-g}  \Big(\zeta(3)^2 e^{-2\phi} + 2 \zeta (3) \zeta (2) + 
\frac{2}{9} \zeta (6) e^{4\phi} \Big) D^6 {\cal{R}}^4\ee
where we have set $ l_s \int d^9 x \sqrt{- g_9} r_B = \int d^{10} x \sqrt{-g}$. We have dropped
a divergent term given by
\be \label{div10} 
\frac{15}{\pi^4} \zeta (5) \zeta (6) l_s^4 \int d^{10} x \sqrt{-g} r_B^4 D^6 {\cal{R}}^4 .\ee
Apart from the genus two term, \C{decomp10} precisely 
matches \C{term10d} providing some more evidence
for the perturbative part of the modular form. Dropping the $f(T, \bar{T}) + f(U,\bar{U})$ 
term in \C{genusval}, note that the ten dimensional contribution comes 
entirely from the terms which are independent of $U$ in \C{genusval}.

Finally, let us consider the divergent term \C{div10}. This has been computed directly in
ten dimensions in~\cite{Green:1999pu}, where it was shown that the divergent term and the genus two
contribution together is proportional to 
\be \frac{2}{3} \zeta (4) e^{4\phi^B} + \frac{1}{2} \zeta (5) r_B^4. \ee 
This is exactly what we get by adding the genus two contribution in \C{decomp10} and the 
divergence in \C{div10}\footnote{We also use $\zeta (4) = \pi^4/90$.}, upto an 
overall irrelevant numerical factor of $\zeta (6)/3 \zeta (4)$. This provides another strong check
of our proposal.

\section{Evidence using eleven dimensional supergravity at one loop on $T^3$}

We now provide some evidence for the perturbative part of the proposed modular form by considering
the four graviton scattering amplitude in eleven dimensional supergravity compactified on $T^3$.
Of course eleven dimensional supergravity cannot give the complete answer. There are extra
contributions due to membrane instantons wrapping the $T^3$. This will give contributions
depending on the Kahler structure modulus in type IIA, and complex structure modulus in type IIB
string theory. So the supergravity analysis will miss such contributions, and we shall see that it 
yields the leading $U_2$ behavior of some of the terms, which arise while going from the M theory
to the string theory coordinates. 

In order to look at the supergravity contributions to the $D^6 {\mathcal{R}}^4$ interaction, 
we need to go beyond the one loop amplitude\footnote{In this section, loops refer to
spacetime loops in eleven dimensional supergravity on $T^3$. We shall refer to the worldsheet expansion 
of string perturbation theory as the genus expansion.}. Two and three loop contributions (and possibly 
higher loops as well) also contribute to the 
amplitude~\cite{Bern:1998ug,Bern:1998zm,Green:1999pu,Bern:2007hh} which we shall not discuss. 
We shall see that the one loop supergravity amplitude coupled with the genus zero string theory 
amplitude will give us some of the terms in our proposed modular form. 

So let us consider one loop supergravity in eleven dimensions compactified on $T^3$. Apart from the overall
kinematic factor which contains the spacetime dependence, the calculation simplifies
and boils down to a box diagram calculation in scalar field theory with cubic interaction, essentially
because of supersymmetry. The four graviton amplitude is given 
by~\cite{Green:1982sw,Russo:1997mk,Green:1997ud,Green:1999pu}    
\be \label{d11}
A_4 = \frac{\kappa_{11}^4}{(2 \pi)^{11}} \hat{K} [I(S,T) + I(S,U) + I (U,T)] , \ee
where $\hat{K}$ involves the ${\cal{R}}^4$ interaction at the linearized level, and 
\be \label{imprel2}
I(S,T) = \frac{2 \pi^4}{l_{11}^3 {V_3}} \int_0^\infty \frac{d \s}{\s} \int_0^1 d 
\omega_3 \int_0^{\omega_3} d \omega_2 \int_0^{\omega_2} d \omega_1 \sum_{\{ l_1,l_2,l_3 \}} 
e^{- G^{IJ} l_I l_J \s/l_{11}^2 -  Q(S,T;\omega_r) \s},\ee 
where $Q(S,T;\omega_r) = -S \omega_1 (\omega_3 - \omega_2) - T (\omega_2 - \omega_1) (1 
- \omega_3)$~\footnote{Note that $\s$ has dimensions of $({\rm length})^2$.}. Here $V_3$ is the
volume of $T^3$ in the M theory metric.
Denoting the torus directions as $1,2$, and $3$, we choose $G_{11} = R_{11}^2$ to be the metric along the 
M theory circle, thus $R_{11} = e^{2\phi^A/3}$.
Though we need the $(s^3 + u^3 + t^3) {\cal{R}}^4$ term, we shall later find it useful to 
extract a part of the momentum independent amplitude from \C{d11} in order
to fix normalizations. This is given by
\bea \label{zeromom}
A_4 (S= T =U =0) &=& \frac{\kappa_{11}^4 \hat{K}}{(2 \pi)^{11}} \cdot 
\frac{\pi^4}{l_{11}^3 {V}_3} \int_0^\infty \frac{d \s}{\s}
\sum_{\{ l_1,l_2,l_3 \}} e^{- G^{IJ} l_I l_J \s/l_{11}^2 } \non \\ &=& \frac{\kappa_{11}^4 
\hat{K}}{(2 \pi)^{11}} \cdot \pi^4 \int_0^\infty \frac{d \s}{\s^{5/2}}
\sum_{\{ \hat{l}_1, \hat{l}_2 , \hat{l}_3 \}} e^{- \frac{\pi G_{IJ} \hat{l}_I \hat{l}_J 
l_{11}^2}{ \s} } ,  \eea
where we have done Poisson resummation using \C{needresum}. Considering the $\hat{l}_1 \neq 0,
\hat{l}_2 = \hat{l}_3 =0$ piece, \C{zeromom} gives~\cite{Green:1997as}
\be \label{partcont1}
A_4 (S= T =U =0) =  \frac{\kappa_{11}^4 \hat{K}}{(2 \pi)^{11} l_{11}^3} \Big[ \pi^3 \zeta (3)
e^{-2\phi^A}+ \ldots \Big] .\ee

Let us now focus on the $(s^3 + u^3 + t^3) {\cal{R}}^4$ interaction, which is
contained in the analytic part of \C{imprel2}. The relevant expression is given by~\cite{Basu:2007ru}
\bea \label{poisson8d}
I (S,T)_{\rm anal} &=& \frac{2\pi^4 {\cal{G}}_{ST}^3}{3! l_{11}^3 V_3} 
\sum_{( l_1,l_2,l_3 ) \neq ( 0,0,0 )} \int_0^\infty d\s \s^2 
e^{- G^{IJ} l_I l_J \s/l_{11}^2} \non \\ &=& \frac{2\pi^7 {\cal{G}}_{ST}^3}{3!}
\sum_{( \hat{l}_1,\hat{l}_2,\hat{l}_3 ) \neq ( 0,0,0 )} \int_0^\infty d\s \sqrt{\s}
e^{- \pi G_{IJ} \hat{l}_I \hat{l}_J l_{11}^2/\s}, \eea
where 
\bea {\cal{G}}_{ST}^3 &=& \int_0^1 d 
\omega_3 \int_0^{\omega_3} d \omega_2 \int_0^{\omega_2} d \omega_1 
\Big( -Q(S,T;\omega_r) \Big)^3 \non \\ &=& \frac{12}{9!} \Big( 
(s^2 t + s t^2) + 3 (s^3 + t^3) \Big) .\eea

We are interested only in those terms in \C{poisson8d} that lead to the perturbative string
contributions given in the previous section. There are two contributions to this:

(i) the $(\hat{l}_2, \hat{l}_3) = (0,0), \hat{l}_1 \neq 0$ part of \C{poisson8d}, which we call 
$I (S,T)_{\rm anal}^1$, and

(ii) the $(\hat{l}_2, \hat{l}_3) \neq (0,0), l_1 =0$ part of \C{poisson8d}, where we have undone the 
Poisson resummation over $\hat{l}_1$ to go to $l_1$, which we call $I (S,T)_{\rm anal}^2$. 

Proceeding
along the lines of~\cite{Basu:2007ru}, we get that
\be \label{partcont2} I (S,T)_{\rm anal}^1 = \frac{\pi^9}{135} {\cal{G}}_{ST}^3 l_{11}^3 e^{2\phi^A} , \ee
where we have used $\zeta (-3) = 1/120$, and
\be \label{partcont3} I (S,T)_{\rm anal}^2 = \frac{2\pi^7 {\cal{G}}_{ST}^3}{3! R_{11} l_{11}} 
\sum_{(\hat{l}_2, \hat{l}_3) \neq (0,0)} \int_0^\infty d\s \s e^{-\pi l_{11}^2 \hat{l}_i \hat{l}_j 
g_{ij}/(\s R_{11})},  \ee 
where we have used the IIA string frame metric
\be g_{i-1,j-1}^A = R_{11} \Big( G_{ij} - \frac{G_{1i} G_{1j}}{G_{11}} \Big),\ee
where $i,j = 2,3$. Using
\be g_{ij}^A = \frac{T_2^A}{U_2^A} \begin{pmatrix} 1 & U_1^A \\ U_1^A & \vert U^A \vert^2 
\end{pmatrix}, \ee 
we get that
\be \label{partcont4} I (S,T)_{\rm anal}^2 = \frac{4 \pi^4}{3!} \Big( \frac{l_{11}}{R_{11}} \Big)^3 
(T_2^A)^2 {\cal{G}}_{ST}^3 E_3 (U^A,\bar{U}^A)^{SL(2,\mathbb{Z})}. \ee

Thus adding \C{partcont2} and \C{partcont4}, we see that the perturbative part is given by
\be I (S,T)_{\rm anal} = \Big[\frac{\pi^9}{135} l_{11}^3 e^{2\phi^A} + \frac{4\pi^4}{3!} (T_2^A)^2 
E_3 (U^A,\bar{U}^A)^{SL(2,\mathbb{Z})} \Big( \frac{l_{11}}{R_{11}} \Big)^3 \Big] {\cal{G}}_{ST}^3.\ee

Finally, using
\be {\cal{G}}_{ST}^3 + {\cal{G}}_{SU}^3 + {\cal{G}}_{UT}^3 = \frac{60}{9!} (s^3 +  t^3 + u^3) ,\ee
we get that
\be \label{partamp}
A_4 =  \frac{\kappa_{11}^4 \hat{K}}{(2 \pi)^{11} l_{11}^3} \Big[ \pi^3 \zeta (3)
e^{-2\phi^A} + \frac{60}{9!} \Big\{ \frac{4 \pi^4}{3!} (T_2^A)^2  
E_3 (U^A,\bar{U}^A)^{SL(2,\mathbb{Z})} + \frac{\pi^9}{135} e^{4\phi^A} \Big\} l_s^6 
(s^3 + t^3 + u^3) \Big],\ee
where we have used $l_{11} = e^{\phi^A/3}l_s$.

In order to fix the genus zero contribution, we note that the tree level amplitude is given by
\be T_2^A e^{-2 \phi^A} \Big( \zeta (3) + \frac{\zeta (3)^2}{2 \cdot 96}  l_s^6 (s^3 + t^3 + u^3) + \ldots \Big)
{\cal{R}}^4 .\ee
Thus given the genus zero ${\cal{R}}^4$ interaction in \C{partamp}, we can also 
deduce the precise coefficient of the $(s^3 + t^3 + u^3) {\cal{R}}^4$ interaction at genus zero. This
contribution has to come from the two loop four graviton amplitude. 

This leads to terms in the IIB effective action in the string frame
\be l_s^6 \int d^8 x \sqrt{-g_8} \Big[ \frac{\pi^3 \zeta (3)^2}{2 \cdot 96} e^{-2\phi} V_2+
\frac{60}{9!} \Big\{ \frac{4 \pi^4}{3!} 
E_3 (T,\bar{T})^{SL(2,\mathbb{Z})} U_2^3 + (e^{-2\phi} V_2 )^{-2} \frac{\pi^9}{135} U_2^3 \Big\} \Big] 
D^6 {\mathcal{R}}^4 . \ee

These are contributions at genus zero, one and three respectively. Given the $U_2$ dependence and the
perturbative equality of the type IIA and type IIB amplitudes, it is natural
to guess that a part of the amplitude with the complete $U$ dependence is
\bea 
\frac{\pi^3 l_s^6}{2 \cdot 96}  \int d^8 x \sqrt{-g_8} \Big[ \zeta (3)^2 e^{-2\phi} V_2+  
\frac{10}{\pi^5} E_3 (T,\bar{T})^{SL(2,\mathbb{Z})} E_3 (U,\bar{U})^{SL(2,\mathbb{Z})}  
\non \\ + \frac{1}{9} (e^{-2\phi} V_2 )^{-2}  
\Big( E_3 (U,\bar{U})^{SL(2,\mathbb{Z})} + E_3 (T,\bar{T})^{SL(2,\mathbb{Z})} 
\Big) \Big] D^6 {\mathcal{R}}^4 , \eea
where we have used $\zeta (6) =\pi^6/945$. This precisely matches some of the terms in \C{genusval}.

\section{The expression for the exact modular form}

Given the expression \C{totcomp} for the perturbative part of the modular form, it is natural to propose
that the exact expression for the modular form is given by
\bea \label{exactcomp}
&&{\mathcal{E}}_{(3/2,3/2)} (M) +\frac{20}{3}
E_{-3/2} (M)^{SL(3,\mathbb{Z})} E_{3} (U, \bar{U})^{SL(2,\mathbb{Z})} \non \\ &&
+ f (U, \bar{U}) +\frac{1}{2}
E_{3/2} (M)^{SL(3,\mathbb{Z})} E_{1} (U, \bar{U})^{SL(2,\mathbb{Z})} ,
\eea
where\footnote{Using \C{relfundantifund}, we could also use the relation $E_{-3/2} (M)^{SL(3,\mathbb{Z})}
= E_{3} (M^{-1})^{SL(3,\mathbb{Z})}$ in \C{exactcomp}.}
\bea \label{pertexact}
{\mathcal{E}}_{(3/2,3/2)} (M)_{\rm pert} &=& \zeta(3)^2 \Big( \tau_2^2 V_2 \Big)^2 + \zeta(3) \tau_2^2 V_2 
E_{1} (T, \bar{T})^{SL(2,\mathbb{Z})}  \non \\ &&+ f (T, \bar{T})
+ \frac{1}{9} \Big( \tau_2^2 V_2 \Big)^{-1}
E_{3} (T, \bar{T})^{SL(2,\mathbb{Z})} . \eea

We now construct $f(T,\bar{T})$ , and also obtain the non--perturbative completion of \C{pertexact}.
Now, the modular form ${\mathcal{E}}_{(3/2,3/2)} (\tau, \bar\tau)$ 
for the $D^6 {\cal{R}}^4$ interaction in ten dimensions satisfies a Poisson equation
\be \label{exp10d}
\Delta_{SL(2,\mathbb{Z})} {\mathcal{E}}_{(3/2,3/2)} (\tau, \bar\tau) = 12 
{\mathcal{E}}_{(3/2,3/2)} (\tau, \bar\tau)
-6 \Big( E_{3/2} (\tau,\bar\tau) \Big)^2\ee
on the fundamental domain of $SL(2,\mathbb{Z})_\tau$~\cite{Green:2005ba}. The source term in \C{exp10d} is the square of 
the modular form for the ${\mathcal{R}}^4$ interaction, which can be understood based on considerations of
supersymmetry. Because $SL(2,\mathbb{Z})_\tau \subset SL(3,\mathbb{Z})_M$, and the $U$ dependence in the expression
\C{exactcomp} is already fixed, it is natural to propose that ${\mathcal{E}}_{(3/2,3/2)} (M)$ satisfies a Poisson 
equation on the fundamental domain of $SL(3,\mathbb{Z})_M$ given by
\be \label{exp8d}
\Delta_{SL(3,\mathbb{Z})} {\mathcal{E}}_{(3/2,3/2)} (M) = \alpha {\mathcal{E}}_{(3/2,3/2)} (M)
+\beta \Big( E_{3/2} (M) \Big)^2,\ee
where $\alpha$ and $\beta$ are numbers.  
Again, the source term in \C{exp8d} is the square of the modular form for the ${\mathcal{R}}^4$ 
interaction in eight dimensions~\cite{Kiritsis:1997em}. 

Let us first consider the perturbative content of \C{exp8d}. 
We use the relation
\be \label{Lappert}
\Delta_{SL(3,\mathbb{Z})}^{\rm pert} =  \Delta_{{SL(2,\mathbb{Z})}_T} + 3 \mu^2 \frac{\p^2}{\p \mu^2}, 
\ee 
where $\mu = \tau_2^2 V_2$ is the eight dimensional dilaton. 
Now \C{Lappert} can be obtained based on symmetries alone. From \C{pertexact}, we see that 
every term in the perturbative part of ${\mathcal{E}}_{(3/2,3/2)} (M)$ is of the form $\mu^k g_k (T,\bar{T})$,
where $g_k (T,\bar{T})$ is $SL(2,\mathbb{Z})_T$ invariant. Thus $\Delta_{SL(3,\mathbb{Z})}^{\rm pert}$
must have the form
\be \Delta_{SL(3,\mathbb{Z})}^{\rm pert} = \xi_1 \Delta_{{SL(2,\mathbb{Z})}_T} +  \xi_2
\mu^2 \frac{\p^2}{\p \mu^2} + \xi_3 \mu \frac{\p}{\p \mu},\ee
where $\xi_1, \xi_2$, and $\xi_3$ are numbers. In order to determine them, we act with 
$\Delta_{SL(3,\mathbb{Z})}^{\rm pert}$ on $E_s (M)_{SL(3,\mathbb{Z})}^{\rm pert}$ which is given by
the first two terms in \C{expEs}, such that $\Delta_{SL(3,\mathbb{Z})}^{\rm pert} 
E_s (M)_{SL(3,\mathbb{Z})}^{\rm pert} = 2s(2s/3 -1) E_s (M)_{SL(3,\mathbb{Z})}^{\rm pert}$. The
first term in \C{expEs} gives $\xi_2 = 3, \xi_3 =0$, while using \C{LapSL2}, we see
that the second term in \C{expEs} gives $\xi_1 =1$, leading to \C{Lappert}.

Using \C{Lappert}, \C{pertexact} and 
\be E_{3/2} (M)_{\rm pert} = 2 \mu \zeta (3) + 2 E_1 (T, \bar{T})^{SL(2,\mathbb{Z})}, \ee
we see that \C{exp8d} gives us the set of equations
\be \label{setofeqns}
\alpha + 4 \beta = 6, \quad \alpha + 8 \beta  =0, \quad \frac{\alpha}{9} = \frac{4}{3}, \ee
and
\be \label{eqnf}
\Delta_{{SL(2,\mathbb{Z})}_T} f (T, \bar{T}) = \alpha f (T, \bar{T}) + 4 \beta
\Big( E_1 (T, \bar{T}) \Big)^2 .\ee
Here we have used the relation \C{LapSL2} for $s=1$ and $s=3$~\footnote{We use the relation 
$\Delta_{{SL(2,\mathbb{Z})}_T} E_1 (T,\bar{T}) =0$ for the unregularized expression.}. 

So \C{setofeqns} is solved by 
\be \alpha = 12, \quad \beta = -\frac{3}{2},\ee
thus \C{eqnf} reduces to
\be \label{eqnfmore} \Delta_{{SL(2,\mathbb{Z})}_T} f (T, \bar{T}) = 12 f (T, \bar{T}) -6
\Big( E_1 (T, \bar{T}) \Big)^2. \ee

Thus \C{eqnfmore} gives us the equation for $f(T,\bar{T})$ (and $f(U,\bar{U})$ as well), 
while \C{exp8d} reduces to
\be \label{simpleqn}
\Delta_{SL(3,\mathbb{Z})} {\mathcal{E}}_{(3/2,3/2)} (M) = 12 {\mathcal{E}}_{(3/2,3/2)} (M)
-\frac{3}{2} \Big( E_{3/2} (M) \Big)^2, \ee
thus giving us an explicit equation satisfied by the modular form ${\mathcal{E}}_{(3/2,3/2)} (M)$. 
Note that the solution of the homogeneous equation $\Delta_{SL(3,\mathbb{Z})} h (M)_{SL(3,\mathbb{Z})} (M) 
= 12 h (M)_{SL(3,\mathbb{Z})} (M)$ (which is the Eisenstein series $E_s (M)_{SL(3,\mathbb{Z})}$ for 
$4s/3 = 1\pm \sqrt{17}$) cannot be added to a particular solution of \C{simpleqn} simply
because this is inconsistent with the structure of terms obtained using string perturbation theory. 

We next understand the structure of $f(T,\bar{T})$ is more detail. 

\subsection{Understanding the structure of $f(T,\bar{T})$}

The structure of \C{eqnfmore} is very similar to \C{exp10d}, which has been analyzed 
in~\cite{Green:2005ba}, and our analysis is along similar lines. In \C{eqnfmore} we substitute 
\be \label{breakf}
f (T,\bar{T}) = f_0 (T_2) + \sum_{k \neq 0} f_k (T_2) e^{2\pi i k T_1} .\ee
Here $f_0 (T_2)$ receives perturbative contributions from the zero worldsheet instanton sector, 
as well as non--perturbative contributions from world sheet instanton and anti--instanton 
pairs of equal and opposite NS--NS charge. 
On the other hand, the remaining part of \C{breakf} receives contributions from world sheet instantons
of non--vanishing NS--NS charge. Substituting the regularized expression for $E_1 (T,\bar{T})$
given by \C{regexp}, we get the 
equation satisfied by $f_0 (T_2)$
\be \label{comp1}
\Big( T_2^2 \frac{\p^2}{\p T_2^2} - 12 \Big) f_0 (T_2) = -6 \Big[ \Big( 2 \zeta (2) T_2 -\pi {\rm ln}
T_2 \Big)^2 + 4 \pi^2  \sum_{k \neq 0}  \mu^2 (k,1) e^{-4\pi \vert k \vert T_2} \Big]. \ee

Now writing 
\be f_0 (T_2) = {\hat{f}}_0 (T_2) + \sum_{k \neq 0} {\hat{f}}_k (T_2) 
e^{-4\pi \vert k \vert T_2}, \ee
where $\hat{f}_0 (T_2)$ is the contribution from the zero worldsheet instanton sector, and 
${\hat{f}}_k (T_2)$ is the contribution from the worldsheet instanton anti--instanton sector
with vanishing NS--NS charge,
from \C{comp1} we get differential equations for ${\hat{f}}_0 (T_2)$ and ${\hat{f}}_k (T_2)$.
For ${\hat{f}}_0 (T_2)$ we get
\be \label{comp3}
\Big( T_2^2 \frac{\p^2}{\p T_2^2} - 12 \Big) \hat{f}_0 (T_2) = -6 \Big( 2 \zeta (2) T_2 -\pi {\rm ln}
T_2 \Big)^2 , \ee
which has the solution
\bea \label{pertsolve}
\hat{f}_0 (T_2) &=& \frac{\pi^2}{720} \Big[ 65 - 20 \pi T_2 + 48 \pi^2 T_2^2 \Big] 
\non \\ &&+ \pi^2 {\rm ln} T_2 \Big[ -\frac{\pi T_2}{3} + \frac{1}{2} {\rm ln} T_2 -\frac{1}{12}\Big]
+ \lambda_1 T_2^4 + \frac{\lambda_2}{T_2^3} ,\eea
where $\lambda_1$ and $\lambda_2$ are arbitrary constants. We shall fix them soon.

For ${\hat{f}}_k (T_2)$,
we get
\be \Big[ T_2^2 \Big( \frac{\p^2}{\p T_2^2} - 8 \pi \vert k \vert \frac{\p}{\p T_2} 
+ (4\pi \vert k \vert)^2 \Big) - 12 \Big) \Big] {\hat{f}}_k (T_2) = - 24 \pi^2 \mu^2 (k,1), \ee
which has the solution
\bea \label{simpnonpert}
{\hat{f}}_k (T_2) &=& -\frac{\mu^2 (k,1)}{448 \vert k \vert^3 \pi T_2^3} 
\Big[ 24 \Big( 4\pi \vert k \vert T_2 + 1 \Big)^2 + \Big( (4\pi \vert k \vert T_2)^3 
-3 \Big)^2 + 15
\non \\ &&+ \Big( 4\pi \vert k \vert T_2 \Big)^4 \Big( 
2 -4\pi \vert k \vert T_2 \Big) 
+ \Big( 4\pi \vert k \vert T_2 \Big)^7 e^{4\pi \vert k \vert T_2} 
{\rm Ei} (-4\pi \vert k \vert T_2)\Big], \eea
where ${\rm Ei}(x)$ is the exponential integral function. Using 
the relation~\cite{GradRyz}
\be {\rm Ei} (-x) = e^{-x} \Big[ -\frac{1}{x} 
+ \int_0^\infty dt \frac{e^{-t}}{(t+x)^2}\Big], \quad x > 0, \ee
we see that the last term in \C{simpnonpert} has the correct structure to be a
worldsheet instanton contribution.

For the worldsheet instantons with non--vanishing NS--NS charge, we get the equation
\bea \label{comp2}
\Big[ T_2^2 \Big( \frac{\p^2}{\p T_2^2} -4\pi^2 k^2 \Big) 
- 12 \Big] f_k (T_2) = -24 \pi \Big( 2 
\zeta (2) T_2 -\pi {\rm ln} T_2 \Big)  \mu (k,1) e^{-2\pi \vert k \vert T_2}\non \\
-24 \pi^2 \sum_{k_1 \neq 0, k_2 \neq 0, k_1 + k_2 = k} \mu (k_1,1) \mu (k_2 ,1) 
e^{-2\pi (\vert k_1 \vert +\vert k_2 \vert) T_2} ,  \eea
which in principle can be solved iteratively by expanding in large $T_2$. 

Substituting \C{pertsolve} and the corresponding expression for $\hat{f}_0 (U_2)$ into \C{exactcomp},
we can easily study the decompactification limit as before. Only the $T_2^2$ term in the
expression for $\hat{f}_0 (T_2)$ (and the $U_2^2$ term in the expression for $\hat{f}_0 (U_2)$)
contributes in this limit. In nine dimensions, in
addition to \C{decomp9} it also gives a term
\be 6 \zeta (4) l_s^5 \int d^9 x \sqrt{-g_9}
(r_B e^{-2 \phi})^{-1} \Big( r_B^2 + \frac{1}{r_B^2} \Big) D^6 {\mathcal{R}}^4 ,\ee
where we have used $\zeta (4) = \pi^4/90$. However, it also gives a divergent contribution 
\be \label{divbad}
\lambda_1 l_s^5 \int d^9 x \sqrt{-g_9} (r_B e^{-2 \phi})^{-1} \Big( r_B^4 
+ \frac{1}{r_B^4} \Big) r_\infty^2 D^6 {\mathcal{R}}^4\ee
which we shall return to soon.

Further decompactifying to ten dimension, this gives an additional contribution to \C{decomp10}
which is equal to
\be 6 \zeta (4) l_s^4 \int d^{10} x \sqrt{-g}  e^{2\phi} D^6 {\cal{R}}^4,\ee
which precisely gives the missing genus two contribution in \C{term10d}. This is a non--trivial
consistency check on our proposed modular form.

Note that we can send 
\be f (T, \bar{T}) \rightarrow f (T,\bar{T}) + \lambda E_4 
(T,\bar{T})^{SL(2,\mathbb{Z})} ,\ee
for arbitrary $\lambda$ in \C{eqnfmore} because $E_4 (T,\bar{T})^{SL(2,\mathbb{Z})}$ satisfies the
homogeneous equation 
\be \label{defE4}
\Delta_{SL(2,\mathbb{Z})} E_4 (T,\bar{T})^{SL(2,\mathbb{Z})} = 12 E_4 
(T,\bar{T})^{SL(2,\mathbb{Z})} .\ee 

In the zero worldsheet instanton sector, this involves shifting the coefficient of the $T_2^4$
term
\be \label{defT4}
\lambda_1 \rightarrow \hat\lambda_1 \equiv \lambda_1 + 2 \lambda \zeta (8),\ee
and the $T_2^{-3}$ term
\be \lambda_2 \rightarrow \hat\lambda_2 \equiv \lambda_2 + \frac{5\pi}{8} 
\lambda \zeta (7) .\ee
In the sector with world sheet instanton charge $k$, the extra terms are 
automatically solutions of the homogeneous equation in \C{comp2}.

We now provide two arguments that we must set the coefficient of the $T_2^4$ term to zero, 
thus $\hat\lambda_1 =0$. From \C{divbad}, note that we get a divergent contribution with a
non--trivial dilaton dependence. As discussed before, the divergences add to give threshold singularities, 
and hence must come only from the genus one amplitude. Thus it follows that $\hat\lambda_1 =0$.
  
The vanishing of $\hat\lambda_1$ can also be argued based on the factorization properties of the amplitude.
Stripping off the eight dimensional dilaton factor from the various loop amplitudes, from 
\C{genusval}, \C{pertsolve}, and \C{defT4}, we see that for large $T_2$, 
the genus two amplitude goes as $T_2^2 + \hat\lambda_1 T_2^4$,
while the genus one amplitude goes as $T_2$. Now considering the degeneration limit of the genus two 
surface into two genus one surfaces as in Figure 2, we see that the large $T_2$ limit of the genus
two amplitude should scale no larger than $T_2^2$, thus $\hat\lambda_1 =0$. 
\begin{figure}[ht]
\[
\mbox{\begin{picture}(300,125)(10,10)
\includegraphics[scale=0.55]{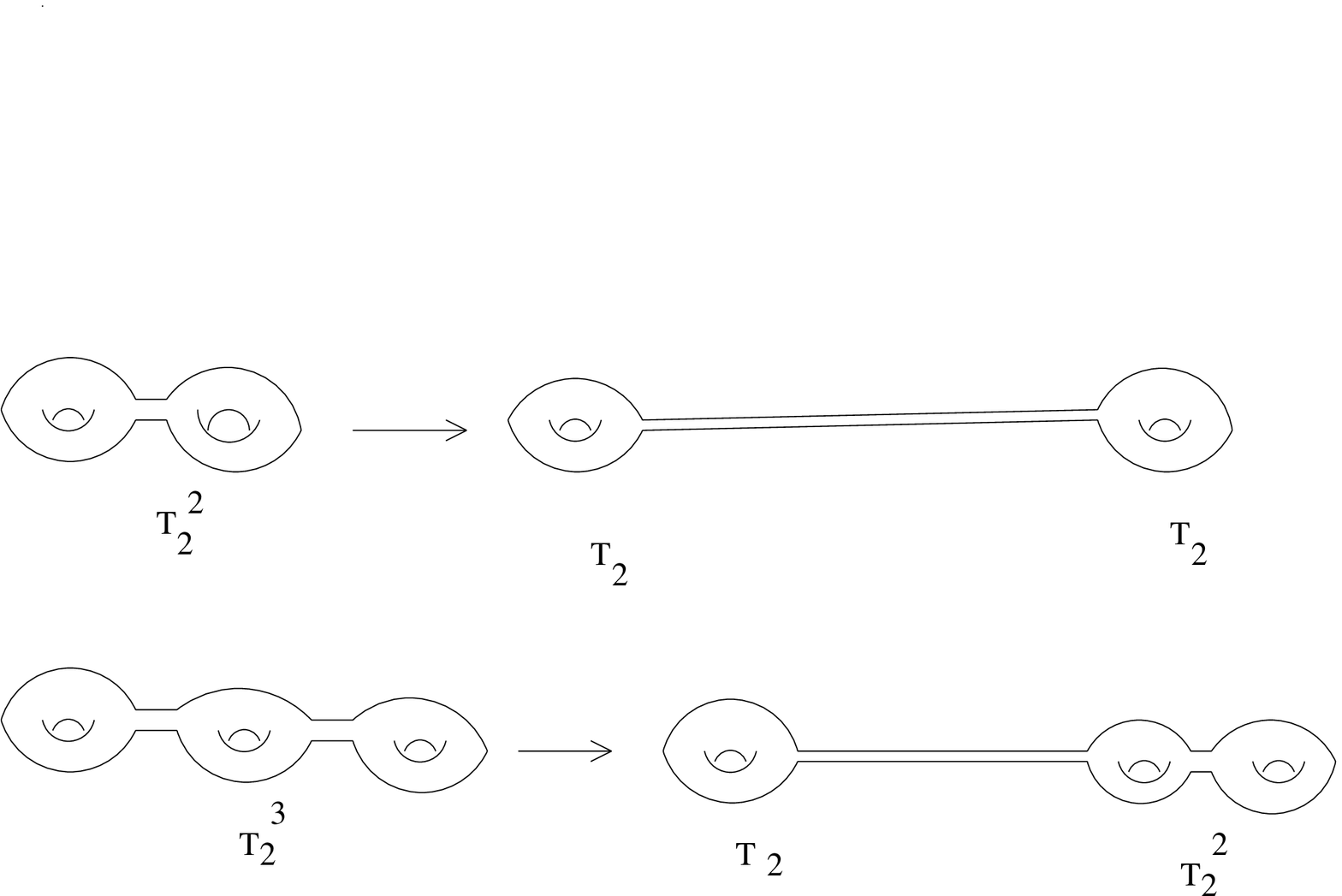}
\end{picture}}
\]
\caption{\it
Degeneration limits of the genus two and genus three surfaces.
}
\end{figure}

Note that from \C{genusval}, it follows that the genus three amplitude at large $T_2$ goes as $T_2^3$.
This is consistent with the degeneration limits described in Figure 2, when $\hat\lambda_1 = 0$. 

We now proceed to calculate $\hat\lambda_2$ along the lines 
of~\cite{Green:2005ba}. Multiplying \C{eqnfmore} by 
$E_4 (T,\bar{T})^{SL(2,\mathbb{Z})}$ and integrating over the restricted fundamental domain of 
$SL(2,\mathbb{Z})_T$, we get that
\bea \label{vallambda2}
\int_{{\mathcal{F}}_L} \frac{d^2 T}{T_2^2} E_4 (T,\bar{T})^{SL(2,\mathbb{Z})} 
\Delta_{{SL(2,\mathbb{Z})}_T} f (T, \bar{T}) = 12 \int_{{\mathcal{F}}_L} \frac{d^2 T}{T_2^2}
E_4 (T,\bar{T})^{SL(2,\mathbb{Z})} f (T, \bar{T}) \non \\
-6 \int_{{\mathcal{F}}_L} \frac{d^2 T}{T_2^2}
E_4 (T,\bar{T})^{SL(2,\mathbb{Z})} \Big( E_1 (T, \bar{T}) \Big)^2. \eea
We have restricted the integral to be over 
${\mathcal{F}}_L$ as the integrals diverge and we regulate them, and finally take
$L \rightarrow \infty$. 
Integrating by parts, and using \C{defE4}, from \C{vallambda2} we get that  
\be \label{needint}
\int_{-1/2}^{1/2} d T_1 \Big( E_4^{SL(2,\mathbb{Z})} \frac{\p f}{\p T_2} - f \frac{ \p 
E_4^{SL(2,\mathbb{Z})}}{\p T_2} \Big)_{T_2 = L} = 
-6 \int_{{\mathcal{F}}_L} \frac{d^2 T}{T_2^2}
E_4 (T,\bar{T})^{SL(2,\mathbb{Z})} \Big( E_1 (T, \bar{T}) \Big)^2. \ee

Using \C{pertsolve} with $\lambda_2$ replaced by $\hat\lambda_2$, the left hand side of 
\C{needint} yields
\be \zeta (8) \Big( -14 \hat\lambda_2 -\frac{4\pi^4}{15} L^5 -\frac{\pi^3}{2} L^4  
-\frac{8\pi^2}{9} L^3 + 2 \pi^3 L^4 {\rm ln} L - 4 \pi^2 L^3 ({\rm ln} L)^2 + \frac{8\pi^2}{3}
L^3 {\rm ln} L \Big). \ee
Using the Poincare series representation for $E_4^{SL(2,\mathbb{Z})}$, and the Rankin--Selberg formula
the right hand side of \C{needint} yields
\bea \zeta (8) \Big( -\frac{48}{5} \zeta(2)^2 L^5 -3\pi \zeta (2) L^4  
-\frac{8\pi^2}{9} L^3 + 12 \zeta (2) L^4 {\rm ln} L - 4 \pi^2 L^3 ({\rm ln} L)^2 + \frac{8\pi^2}{3}
L^3 {\rm ln} L \Big) \non \\
-48 \pi^2 \zeta (8)  \int_0^L d T_2 T_2^2 \sum_{k \neq 0} \mu^2 (k,1) 
e^{-4 \pi \vert k \vert T_2} ,\eea
leading to
\be \hat\lambda_2 = \frac{3}{14\pi} \sum_{k=1}^\infty \frac{\mu^2 (k,1)}{k^3} 
= \frac{1}{4} \zeta (3) \zeta (5),\ee
using an identity due to Ramanujan~\cite{Apostol}.

\subsection{Understanding the non--perturbative structure of ${\mathcal{E}}_{(3/2,3/2)} (M)$}

Having understood the perturbative part of ${\mathcal{E}}_{(3/2,3/2)} (M)$, let us focus on the
non--perturbative part of ${\mathcal{E}}_{(3/2,3/2)} (M)$. From \C{simpleqn}, we can see what
are the various kinds of non--perturbative contributions ${\mathcal{E}}_{(3/2,3/2)} (M)$
receives. This allows us to write
\bea \label{contnp}
{\mathcal{E}}_{(3/2,3/2)} (M)_{\rm non-pert} =  \sum_{k \neq 0} ( f_k (\phi_i) e^{2\pi i k \tau_1}
+ u_k (\phi_i) ) +\sum_{k \neq 0} ( g_k (\phi_i,\tau_1) e^{2\pi i k B_R} + v_k (\phi_i,\tau_1) ) \non \\
+\sum_{k \neq 0, l \neq 0} h_{k,l} (\phi_i,\tau_1) e^{2\pi i(k \tau_1
+ l B_R)}  ,\eea
where $\phi_i = \{ B_N, V_2, \tau_2 \}$. In \C{contnp}, $ f_k (\phi_i)$ involves 
charge $k$ (single and double) D--instanton contributions, while $g_k (\phi_i,\tau_1)$ involves 
(single and double) $(p,q)$ string instanton
contributions carrying R--R charge $k$. The $h_{k,l} (\phi_i,\tau_1)$ term involves contributions
from charge $k$ D--instantons and R--R charge $l$ $(p,q)$ string instantons put together.
Also $u_k (\phi_i)$ includes D--instanton anti--D--instanton contributions with total charge zero, 
which goes as $e^{-4\pi \vert k \vert \tau_2}$ for large $\tau_2$.
Finally $v_k (\phi_i,\tau_1 )$ includes $(p,q)$ and $(p',q')$ string instanton contributions with total
R--R charge zero, which goes as $e^{-4\pi \vert k \tau \vert V_2}$ in the sector with only D--strings.

From \C{simpleqn}, we obtain explicit differential equations satisfied 
by these non--perturbative contributions. Defining 
\be \hat\Delta = \tau_2^2 \frac{\p^2}{\p \tau_2^2} + V_2^2 \p_{B_N}^2 + 3 \p_\nu
(\nu^2 \p_\nu), \ee
we get that
\bea &&\Big( \hat\Delta -4\pi^2 k^2 \tau_2^2 -12\Big) f_k (\phi_i) \non \\ &&= 
-48 \pi \tau_2 V_2 \Big(  \tau_2^2 V_2 \zeta (3) +  E_1 (T, \bar{T})^{SL(2,\mathbb{Z})} \Big)
\vert k \vert \mu (k,\frac{3}{2}) K_1 (2\pi \vert k \vert \tau_2) \non \\ 
&&- 96 (\pi \tau_2 V_2)^2
\sum_{k_i \neq 0 , k_1 + k_2 = k} \vert k_1 k_2 \vert \mu (k_1,\frac{3}{2}) \mu (k_2,\frac{3}{2})
K_1 (2\pi \vert k_1 \vert \tau_2) K_1 (2\pi \vert k_2 \vert \tau_2). \eea 
Further defining
\be \mu (k,l, s) = \sum_{m > 0, m | k,l} \frac{1}{m^{2s -1}},\ee
such that $\mu (k,0,s)  = \mu (k,s)$, we also get that
\bea
&&\Big( \hat\Delta + \tau_2^2 \p_{\tau_1}^2
- V_2^2 [4\pi^2 k^2 \vert \tau \vert^2 + 4\pi i k \tau_1 \p_{B_N}] -12\Big) 
g_k (\phi_i,\tau_1) =\non \\ &&-24 \pi\Big(  \tau_2^2 V_2 \zeta (3) +  E_1 (T, \bar{T})^{SL(2,\mathbb{Z})} \Big)
\sum_l \mu (k,l,1) e^{-2\pi \vert l -k \tau \vert V_2 + 2\pi i l B_N} \non \\
&&-24 \pi^2 \sum_{k_i \neq 0, l_i, k_1 + k_2 = k} \mu (k_1,l_1,1) \mu (k_2,l_2,1) e^{-2\pi (\vert l_1 - k_1 
\tau \vert + \vert l_2 - k_2 \tau \vert )V_2 + 2\pi i (l_1 + l_2) B_N },\eea
and
\bea
&&\Big( \hat\Delta + \tau_2^2 [\p_{\tau_1}^2 + 4\pi i k \p_{\tau_1}
- 4\pi^2 k^2] - V_2^2 [4\pi^2 k^2 \vert \tau \vert^2 + 4\pi i k \tau_1 \p_{B_N}] -12\Big) 
h_{k,l} (\phi_i,\tau_1) = \non \\ &&-96 \pi^2 \tau_2 V_2 \sum_m \vert k \vert \mu (k,\frac{3}{2}) \mu (l,m,1)
K_1 (2\pi \vert k \vert \tau_2) e^{-2\pi \vert m - l \tau \vert V_2 + 2\pi i m B_N}.\eea
The remaining two differential equations are given by
\bea
&&\Big( \hat\Delta -12\Big) u_k (\phi_i) = - 96 (\pi \tau_2 V_2)^2 \vert k \vert^2 \mu^2 (k,\frac{3}{2})
K_1^2 (2\pi \vert k \vert \tau_2), \eea
and
\bea
&&\Big( \hat\Delta + \tau_2^2 \p_{\tau_1}^2 -12 \Big) v_k (\phi_i,\tau_1) 
\non \\ &&= -24 \pi^2 \sum_{l_1, l_2} \mu (k, l_1, 1) \mu (k,l_2, 1)
e^{-2\pi (\vert l_1 -  k \tau \vert + \vert l_2 + k \tau \vert) V_2 + 2\pi i (l_1 + l_2) B_N}.
\eea

\section{More predictions from eleven dimensional supergravity at one loop on $T^3$}

We can generalize the calculations in section 3 to make predictions for some of the perturbative
contributions to the $D^{2k} {\cal{R}}^4$ interaction for arbitrary values of $k \geq 4$. We show below
that we obtain parts of the genus one and genus $k$ contributions to the amplitude. However, 
it need not be the case that the $D^{2k} {\cal{R}}^4$ interaction is protected for all values of 
$k$.
 
The analytic part of the amplitude relevant for the $D^{2k} {\cal{R}}^4$ interaction 
is given by~\cite{Basu:2007ru}
\bea 
I (S,T)_{\rm anal} &=& \frac{2\pi^4 {\cal{G}}_{ST}^k}{k! l_{11}^3 V_3} 
\sum_{( l_1,l_2,l_3 ) \neq ( 0,0,0 )} \int_0^\infty d\s \s^{k-1} 
e^{- G^{IJ} l_I l_J \s/l_{11}^2} \non \\ &=& \frac{2\pi^{4+k} {\cal{G}}_{ST}^k}{k!}
\sum_{( \hat{l}_1,\hat{l}_2,\hat{l}_3 ) \neq ( 0,0,0 )} \int_0^\infty d\s \s^{k-5/2}
e^{- \pi G_{IJ} \hat{l}_I \hat{l}_J l_{11}^2/\s}, \eea
where 
\bea {\cal{G}}_{ST}^k = \int_0^1 d 
\omega_3 \int_0^{\omega_3} d \omega_2 \int_0^{\omega_2} d \omega_1 
\Big( -Q(S,T;\omega_r) \Big)^k .\eea

Following the same steps as in section 3, the two perturbative contributions are given by
\be I (S,T)_{\rm anal}^1 = 4\pi^{2k + 5/2} \Gamma \Big(\frac{3}{2} -k \Big) \zeta (3-2k) l_{11}^{2k-3}
e^{2(2k-3)\phi^A/3} \frac{{\cal{G}}_{ST}^k}{k!} , \ee
and
\be I (S,T)_{\rm anal}^2 = \frac{2 \pi^4 l_{11}^{2k-3}}{k R_{11}^k}  
(T_2^A)^{k-1} {\cal{G}}_{ST}^k E_k (U^A,\bar{U}^A)^{SL(2,\mathbb{Z})} .\ee
This leads to
\bea \label{highk}
A_4 &=&  \frac{\kappa_{11}^4 \hat{K}}{(2 \pi)^{11} l_{11}^3} \Big[ \frac{2 \pi^4}{k} 
(T_2^A)^{k-1} E_k (U^A,\bar{U}^A)^{SL(2,\mathbb{Z})} \non \\
&& + \frac{4\pi^{2k + 5/2}}{k!} 
\Gamma \Big(\frac{3}{2} -k \Big) \zeta (3-2k) e^{2(k-1)\phi^A} \Big] l_s^{2k} {\mathcal{W}}^k ,\eea
where
\be {\mathcal{W}}^k = {\cal{G}}_{ST}^k + {\cal{G}}_{SU}^k + {\cal{G}}_{UT}^k. \ee
Now ${\mathcal{W}}^k$ contains all the possible $2k$-th power of the derivatives acting on 
${\cal{R}}^4$ consistent with the kinematical structure of the amplitude. This is unique upto $k=5$, 
namely, for $k=4$, ${\mathcal{W}}^4 \sim (s^2 + t^2 + u^2)^2$, while for $k=5$, 
${\mathcal{W}}^5 \sim (s^2 + t^2 + u^2) (s^3 + t^3 + u^3)$.  For $k=6$, there are two 
independent structures
and so ${\mathcal{W}}^6 \sim (s^2 + t^2 + u^2)^3 + (s^3 + t^3 + u^3)^2$, leading to two different
spacetime structures for the $D^{12} {\cal{R}}^4$ interaction. Thus when we mean the $D^{2k} {\cal{R}}^4$ 
interaction, we mean that these various possibilities have already been taken into account. 
Thus, \C{highk} leads to terms in the IIB effective action given by
\bea \label{termskact}
&&l_s^{2k} \int d^8 x \sqrt{-g_8} \Big[ \frac{2 \pi}{k} 
(U_2^B)^k E_k (T^B,\bar{T}^B)^{SL(2,\mathbb{Z})} \non \\ &&~~~~~~~~~~~~~~~~~~+ \frac{4\pi^{2k - 1/2}}{k!} 
\Gamma \Big(\frac{3}{2} -k \Big) \zeta (3-2k) (e^{-2\phi^B} T_2^B)^{1-k} (U_2^B)^k\Big] 
D^{2k} {\mathcal{R}}^4 . \eea
Given the perturbative equality of the amplitude in the two type II theories, and \C{expSL2}, it is natural
to enhance the $(U_2^B)^k$ factors to $E_k (U^B,\bar{U}^B)^{SL(2,\mathbb{Z})}$, and symmetrize in $U^B$ and
$T^B$. Thus \C{termskact} gets enhanced to
\bea \label{termskact2}
&&l_s^{2k} \int d^8 x \sqrt{-g_8} \Big[ \frac{(2k)!}{(2\pi)^{2k-1} \vert B_{2k} \vert k } 
E_k (T^B,\bar{T}^B)^{SL(2,\mathbb{Z})} E_k (U^B,\bar{U}^B)^{SL(2,\mathbb{Z})} \non \\ &&+ 
\frac{4 \Gamma (k + \frac{1}{2}) \Gamma (k-1) \zeta (2k -2)}{\pi^{2k -3/2} 
\vert B_{2k} \vert} (e^{-2\phi^B} T_2^B)^{1-k} \non \\ &&\times
\Big( E_k (T^B,\bar{T}^B)^{SL(2,\mathbb{Z})} + E_k (U^B,\bar{U}^B)^{SL(2,\mathbb{Z})} \Big)\Big] 
D^{2k} {\mathcal{R}}^4 , \eea
where we have used the relations~\cite{GradRyz}
\be \zeta (2k) = \frac{2^{2k-1} \pi^{2k} \vert B_{2k} \vert}{(2k)!},\ee
where $k$ is a positive integer, $B_{2k}$ are the Bernoulli numbers, the identity \C{imprel}, and 
\be \Gamma (2 x) = \frac{2^{2x - 1/2}}{\sqrt{2\pi}} \Gamma (x) \Gamma (x+ \frac{1}{2}) .\ee
Thus from \C{termskact2}, we see that eleven dimensional supergravity gives predictions for 
parts of the genus one and genus $k$ amplitudes
for the $D^{2k} {\mathcal{R}}^4$ interaction, for arbitrary $k$. 

Thus from \C{termskact2}, we see that at genus one, there is a contribution 
proportional to $E_k (T^B,\bar{T}^B)^{SL(2,\mathbb{Z})} E_k (U^B,\bar{U}^B)^{SL(2,\mathbb{Z})}$.
For low values of $k$, it is easy to see that there is such a contribution. For $k=2$,
as shown in~\cite{Basu:2007ru}, this arises from the only diagram that contributes 
to the torus amplitude given by figure 3.  

\begin{figure}[ht]
\[
\mbox{\begin{picture}(95,65)(10,10)
\includegraphics[scale=0.30]{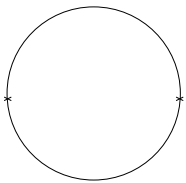}
\end{picture}}
\]
\caption{\it
Schematics of the $D^4 R^4$ torus amplitude.
}
\end{figure}

For $k=3$, from figure 1, we see that $\hat{I}_1$ gives such a contribution proportional to
$E_3 (T^B,\bar{T}^B)^{SL(2,\mathbb{Z})} E_3 (U^B,\bar{U}^B)^{SL(2,\mathbb{Z})}$. However, there
is also another contribution from $\hat{I}_2$. 

\begin{figure}[ht]
\[
\mbox{\begin{picture}(95,65)(10,10)
\includegraphics[scale=0.40]{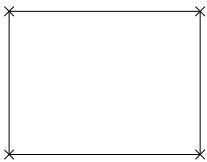}
\end{picture}}
\]
\caption{\it
Schematics of part of the $D^8 R^4$ torus amplitude.
}
\end{figure}

For $k=4$, again we can see that the part of the torus
amplitude coming from the diagram in figure 4 is proportional to
$E_4 (T^B,\bar{T}^B)^{SL(2,\mathbb{Z})} E_4 (U^B,\bar{U}^B)^{SL(2,\mathbb{Z})}$. This can be obtained
by using the relation
\bea \int_{\cal{T}} \prod_{i=1}^3 \frac{d^2 
\nu_i}{\Omega_2}   {\rm ln } \hat\chi (\nu_1 - \nu_2 ; 
\Omega) {\rm ln } \hat\chi (\nu_1 - \nu_3 ;\Omega) \hat\chi (\nu_2 - \nu_4 ;\Omega) 
\hat\chi (\nu_3 - \nu_4 ;\Omega)\non \\
= \frac{1}{(4 \pi)^4} \sum_{(m,n) \neq (0,0)} \frac{\Omega_2^4}{\vert m \Omega + n \vert^8}
= \frac{1}{(4 \pi)^4} E_4
(\Omega, \bar\Omega)^{SL(2,\mathbb{Z})} ,\eea
and generalizing the calculation of $\hat{I}_1$ summarised in Appendix B.1. However, just like 
in the $k=3$ case, other parts of the torus amplitude should also give the same contribution,
so the final numerical coefficient will be different. 

So from the discussion above, one can see that when the $k$ points form a polygon
with no internal lines, the integral over the vertex operator insertions is
proportional to $ E_k (\Omega, \bar\Omega)^{SL(2,\mathbb{Z})}$, while leads to the
contribution predicted from supergravity. However, this topology is no more possible
for $k \geq 5$, and so there is no particularly simple contribution to the torus 
amplitude that gives the answer. The various contributions must add to give
the answer predicted from supergravity. It would be interesting to see this 
explicitly coming out of the torus amplitude.

After converting to the Einstein frame, let us consider the $U^B$ dependent coefficient of 
the $\hat{D}^{2k} \hat{\mathcal{R}}^4$ interaction in \C{termskact2}.\footnote{The 
remaining part which depends only on $T^B$ must form part of an $SL(3,\mathbb{Z})_M$
invariant modular form.} Since it involves
$E_k (U^B,\bar{U}^B)^{SL(2,\mathbb{Z})}$ which is $SL(2,\mathbb{Z})_U$ invariant, whatever
multiplies it in the whole amplitude should be $SL(3,\mathbb{Z})_M$ invariant. In fact, this 
contribution is given by
\bea &&\frac{2\Gamma (k + \frac{1}{2})}{\pi \vert B_{2k} \vert } \Big(
2\pi^{5/2 -2 k} (e^{-2\phi^B} T_2^B)^{1-2k/3} \Gamma (k-1) \zeta (2k-2)
\non \\ &&+ \pi^{3/2-2k} \Gamma (k) (e^{-2\phi^B} T_2^B)^{k/3} 
E_k (T^B,\bar{T}^B)^{SL(2,\mathbb{Z})}\Big) E_k (U^B,\bar{U}^B)^{SL(2,\mathbb{Z})} 
\non \\ &&= \frac{2\Gamma (k + \frac{1}{2}) \Gamma (\frac{3}{2} -k)}{\pi \vert B_{2k} \vert }
E_{3/2 -k} (M)^{SL(3,\mathbb{Z})}_{\rm pert} E_k (U^B,\bar{U}^B)^{SL(2,\mathbb{Z})} \non \\
&&= \frac{2\Gamma (k + \frac{1}{2}) \Gamma (\frac{3}{2} -k)}{\pi \vert B_{2k} \vert }
E_{k} (M^{-1})^{SL(3,\mathbb{Z})}_{\rm pert} E_k (U^B,\bar{U}^B)^{SL(2,\mathbb{Z})}, \eea
on using \C{relfundantifund}. Extending it to the non--perturbative completion, we get
the manifestly U--duality invariant modular form
\be \frac{2\Gamma (k + \frac{1}{2}) \Gamma (\frac{3}{2} -k)}{\pi \vert B_{2k} \vert }
E_{k} (M^{-1})^{SL(3,\mathbb{Z})} E_k (U^B,\bar{U}^B)^{SL(2,\mathbb{Z})}.\ee 
Thus one loop supergravity and U--duality gives a prediction for a part of the 
complete modular form.
 
Decompactifying to nine dimensions, we see that \C{termskact2} gives the interaction
\bea  \label{termskact9}
&&l_s^{2k -1} \int d^9 x \sqrt{-g_9} \Big[ \frac{4\pi^{3/2}}{k!} \zeta (2k-1) \Gamma \Big(k- 
\frac{1}{2} \Big) \Big( r_B^{2k-1} + \frac{1}{r_B^{2k-1}} \Big) \non \\ &&+ 
4\pi^2 \frac{\zeta (2k-2)}{k(k-1)} (e^{-2\phi^B} r_B)^{1-k} 
\Big( r_B^{k} + \frac{1}{r_B^{k}} \Big)\Big] 
D^{2k} {\mathcal{R}}^4 , \eea
which contributes at genus one and at genus $k$. It also gives the divergent contribution
\be \frac{4\pi}{k} \zeta (2k) l_s^{2k -1} \int d^9 x \sqrt{-g_9} 
r_\infty^{2k-1} D^{2k} {\mathcal{R}}^4 \ee
which leads to the threshold singularities. Further decompactifying \C{termskact9} to ten dimensions, this leads 
to the interaction
\be 4\pi^2 \frac{\zeta (2k-2)}{k(k-1)} l_s^{2k -2} \int d^{10} x \sqrt{-g} e^{-2(1-k)\phi^B} 
D^{2k} {\mathcal{R}}^4 ,\ee
which contributes at genus $k$, while the genus one contribution vanishes. It also gives the
divergent contribution
\be \frac{4\pi^{3/2}}{k!} \zeta (2k-1) \Gamma \Big(k- 
\frac{1}{2} \Big) l_s^{2k -2} \int d^{10} x \sqrt{-g} r_B^{2(k-1)} D^{2k} {\mathcal{R}}^4, \ee
corresponding to the threshold singularities.

\section{Discussion}

We have made a proposal for the modular form for the $D^6 {\cal{R}}^4$ 
interaction, and showed that is satisfies several non--trivial consistency checks.
Some parts of the torus amplitude, however, have been constructed based on the
perturbative equality of the type IIA and type IIB amplitudes, and some heuristic arguments.
Calculating the full amplitude explicitly would be useful in verifying the proposal we make. 

Let us make some comments about the possible modular form for the $D^6 {\cal{R}}^4$ interaction
in toroidal compactifications preserving maximal supersymmetry to
lower dimensions, where the U--duality group is no longer reducible.
The scalars parametrize the coset manifold ${\cal{M}} = G/H$, where $G$ is a 
non--compact group, and $H$ is its maximal compact subgroup~\cite{Cremmer:1980gs,Julia:1980gr}. 
The conjectured U--duality group is $\hat{G}$, the discrete version of $G$. Thus in the Einstein 
frame the term in the supergravity action involving the scalars is given by 
\be S \sim \frac{1}{l_s^{8 -d}} \int d^{10-d} x \sqrt{-\hat{g}_{10-d}}  
{\rm Tr} (\p_\mu M \hat{\p}^\mu M^{-1}) ,\ee
where $M$ parametrizes ${\cal{M}}$. Based on the $D^6 {\cal{R}}^4$ interaction in ten
dimensions as well as the modular form we propose, it is conceivable that the 
U--duality invariant modular form in lower dimensions is given by the solution of the Poisson 
equation on the fundamental domain of ${\hat{G}}$ given by
\be \Delta_{\hat{G}} {\mathcal{E}}_{(3/2,3/2)} (M) = \lambda_1 
{\mathcal{E}}_{(3/2,3/2)} (M) -\lambda_2 \Big( E_{3/2}  (M) \Big)^2 ,\ee
where $\lambda_1$ and $\lambda_2$ are constants.

\section*{Acknowledgements}

I would like to thank Michael Green and Pierre Vanhove for useful comments.
The work of A.~B. is supported by NSF Grant No.~PHY-0503584 and the William D. Loughlin membership.

\section{Appendix}

\appendix

\section{Expressions for the Eisenstein series}

In the section below, we write down explicit expressions for the Eisenstein series of 
$SL(2,\mathbb{Z})$ and $SL(3,\mathbb{Z})$ that are useful in the main text.

\subsection{The Eisenstein series for $SL(2,\mathbb{Z})$}

The Eisenstein series of order $s$ for $SL(2,\mathbb{Z})$ 
is defined by
\bea \label{expSL2}
E_{s} (T,\bar{T})^{SL(2,\mathbb{Z})}  &=& \sum_{(p,q) \neq (0,0)} 
\frac{T_2^{s}}{\vert p + q T\vert^{2s }} \non \\
&=& 2 \zeta(2s) T_2^s + 2 \sqrt{\pi} T_2^{1-s} \frac{\Gamma (s -1/2)}{\Gamma (s)}
\zeta (2s -1) \non \\ &&+ \frac{2 \pi^s \sqrt{T_2}}{\Gamma (s)} \sum_{m_1 \neq 0 ,m_2 \neq 0} 
\Big\vert \frac{m_1}{m_2} \Big\vert^{s - 1/2} K_{s - 1/2} (2\pi T_2 \vert m_1 m_2
\vert) e^{2\pi i m_1 m_2 T_1} \non \\ 
&=& 2 \zeta(2s) T_2^s + 2 \sqrt{\pi} T_2^{1-s} \frac{\Gamma (s -1/2)}{\Gamma (s)}
\zeta (2s -1) \non \\ &&+ \frac{4 \pi^s \sqrt{T_2}}{\Gamma (s)} \sum_{k \neq 0} 
\vert k \vert^{s - 1/2} \mu (k,s) K_{s - 1/2} (2\pi T_2 \vert k
\vert) e^{2\pi i k T_1}, \eea
where
\be \mu (k,s) = \sum_{m > 0, m | k} \frac{1}{m^{2s -1}}.\ee

Using the relations

\be \label{imprel}
\zeta (2s -1 ) \Gamma (s - \frac{1}{2}) = \pi^{2s - 3/2} \zeta (2 - 2s) \Gamma (1-s) , \ee
and 
\be K_s (x) = K_{-s} (x) , \ee
we see that
\be \label{relSL2}
\Gamma (s) E_{s} (T,\bar{T})^{SL(2,\mathbb{Z})}  = \pi^{2s - 1} \Gamma (1-s) 
E_{1-s} (T,\bar{T})^{SL(2,\mathbb{Z})} . \ee

Now \C{expSL2} satisfies the Laplace equation 
\be \label{LapSL2}
\Delta_{SL(2,\mathbb{Z})} E_s (T,\bar{T})^{SL(2,\mathbb{Z})} = 4 T_2^2 \frac{\p^2}{\p T \p \bar{T}} 
E_s (T,\bar{T})^{SL(2,\mathbb{Z})} = s (s -1) E_s (T,\bar{T})^{SL(2,\mathbb{Z})}\ee
on the fundamental domain of $SL(2,\mathbb{Z})_T$.

We shall need the expression for $E_1 (T,\bar{T})^{SL(2,\mathbb{Z})}$ in the main text. From \C{expSL2},
note that this diverges because $\zeta (1)$ is infinite, and thus needs to be regularized. We regularize
the second term in \C{expSL2} by setting $1-s = \epsilon$ and taking the limit $\epsilon \rightarrow 0$,
where we also use
\be \zeta (\epsilon) = \frac{1}{\epsilon} +\gamma + O (\epsilon),\ee
where $\gamma$ is the Euler constant.
Using an $\overline{MS}$--like regularization scheme, where we drop the $1/\epsilon$ pole term as well terms 
involving the Euler constant, we get that (using $\zeta (2) = \pi^2/6$)
\bea \label{regexp}
E_1 (T,\bar{T})^{SL(2,\mathbb{Z})} &=& \frac{\pi^2}{3} T_2 -\pi {\rm ln} T_2
\non \\ &&+ 2 \pi \sqrt{T_2} \sum_{m \neq 0 ,n \neq 0} 
\Big\vert \frac{m}{n} \Big\vert^{1/2} K_{1/2} (2\pi T_2 \vert m n
\vert) e^{2\pi i m n T_1} \non \\ &=& - \pi {\rm ln} \Big( T_2 \vert \eta (T) \vert ^4 \Big), \eea
where we have used
\be K_{1/2} (x) = \sqrt{\frac{\pi}{2x}}e^{-x} ,\ee
and the definition of the Dedekind eta function
\be \eta (T) = e^{\pi i T/12} \prod_{k =1}^\infty (1- e^{2\pi i k T}). \ee
This yields the same result as in~\cite{Dixon:1990pc}.

\subsection{The Eisenstein series for $SL(3,\mathbb{Z})$}

The Eisenstein series of order $s$ for $SL(3,\mathbb{Z})$ in the fundamental representation
is defined by
\bea  \label{expSL3}
E_s (M)^{SL(3,\mathbb{Z})}  &=& \sum'_{m_i} \Big( m_i M_{ij} m_j \Big)^{-s} \non \\
& =&\sum'_{m_i} \nu^{-s/3} \Big( \frac{\vert m_1 + m_2 \tau + m_3 B
\vert^2}{\tau_2}+\frac{m_3^2}{\nu} \Big)^{-s}, \eea
where $m_i$ are integers, and the sum excludes $\{ m_1, m_2, m_3 \} = \{ 0,0,0 \}$.
The integers $m_i$ transform in the anti--fundamental representation of $SL(3,\mathbb{Z})$, and 
the matrix $M_{ij}$ is given by \C{matrix}.

Using the integral representation
\bea  \label{expE}
E_s (M)^{SL(3,\mathbb{Z})}  =
\frac{\nu^{-s/3} \pi^s}{\Gamma (s)} \int_0^\infty \frac{d t}{t^{s + 1}} \sum'_{m_i} 
e^{-\pi ( \vert m_1 + m_2 \tau + m_3 B
\vert^2/\tau_2+ m_3^2/\nu )/t }, \eea
we can evaluate \C{expE} to get that
\bea \label{expEs}
&&E_s (M)^{SL(3,\mathbb{Z})} = 2 ( \tau_2^2 V_2)^{2s/3} \zeta (2s) + \frac{\sqrt{\pi} \Gamma (s -  
1/2)}{\Gamma (s)}
(\tau_2^2 V_2)^{1/2 - s/3} E_{s -1/2} (T,\bar{T})^{SL(2,\mathbb{Z})} \non \\&&
+ \frac{2 \pi^s}{\Gamma(s)} \tau_2^{s/3 + 1/2} V_2^{2s/3} 
\sum_{m_1 \neq 0 ,m_2 \neq 0} 
\Big\vert \frac{m_1}{m_2} \Big\vert^{s - 1/2} K_{s - 1/2} (2\pi \tau_2 \vert m_1 m_2
\vert) e^{2\pi i m_1 m_2 \tau_1} \non \\  &&+ \frac{2 \pi^s}{\Gamma(s)} \tau_2^{1 -2s/3} V_2^{1 - s/3} 
\sum_{m_1 \neq 0, m_3 \neq 0, m_2} \Big\vert \frac{m_2 - m_1 \tau}{m_3} 
\Big\vert^{s-1} K_{s-1} (2\pi \vert m_3
( m_2 - m_1 \tau) \vert V_2) \non \\ && ~~~~~~~~~~~~~~~~~~~~~~~~~~~~~~~~
~~~~~~~~~~~~~~~~~~~~~~~~~~~~~~~~~~~~~\times e^{2\pi i m_3 (m_1 B_R + m_2 B_N)}. \eea

Now \C{expSL3} satisfies the Laplace equation~\cite{Kiritsis:1997em}  
\bea \Delta_{SL(3,\mathbb{Z})} E_s (M)^{SL(3,\mathbb{Z})} &=& 
\Big[ 4 \tau_2^2 \frac{\p^2}{\p \tau \p \bar\tau} 
+ \frac{1}{\nu \tau_2} \vert \p_{B_N} - \tau \p_{B_R} \vert^2 + 3 \p_\nu 
( \nu^2 \p_\nu ) \Big]E_s (M)^{SL(3,\mathbb{Z})} 
\non \\ &&= \frac{2s (2s -3)}{3} E_s (M)^{SL(3,\mathbb{Z})} \eea
on the fundamental domain of $SL(3,\mathbb{Z})_M$.

We can also define the Eisenstein series of order $s$ in the anti--fundamental representation
by
\bea  \label{expEanti}
E_s (M^{-1})^{SL(3,\mathbb{Z})}  = \sum'_{\hat{m}_i} \Big( \hat{m}_i M^{ij} \hat{m}_j \Big)^{-s}, \eea
where $\hat{m}_i$ transforms in the fundamental representation of $SL(3,\mathbb{Z})$. Now using the 
result 
\be \label{needresum}
\sum'_{\hat{l}_i} e^{-\pi \s G^{ij} \hat{l}_i \hat{l}_j} = \s^{-3/2} \sqrt{{\rm det} G} \sum'_{l_i} 
e^{-\pi G_{ij} l_i l_j/\s} \ee
for invertible matrices, which can be derived using Poisson resummation, we get that
\be \label{relfundantifund}
E_s (M^{-1})^{SL(3,\mathbb{Z})}  = E_{3/2 -s} (M)^{SL(3,\mathbb{Z})}  . \ee

Thus there is a simple relationship between the Eisenstein series for the fundamental and the
anti--fundamental representations.

\section{Calculating $\hat{I}_1$ and $\hat{I}_2$}

Here we provide various details of calculating $\hat{I}_1$ and $\hat{I}_2$ which are needed to calculate the
torus amplitude.

\subsection{Calculating $\hat{I}_1$}

We first evaluate \C{torone}, for which we use the representation
\be \label{proptorus}
{\rm ln } \hat\chi (\nu ; \Omega) = \frac{1}{4\pi} \sum_{(m,n) \neq (0,0)} \frac{\Omega_2}{\vert m \Omega 
+ n \vert^2} e^{\pi [ \bar\nu (m\Omega + n )-\nu(m\bar\Omega + n)]/\Omega_2} \ee
for the scalar propagator on the torus. This leads to the relation~\cite{Green:1999pv}
\bea \int_{\cal{T}} \prod_{i=1}^3 \frac{d^2 
\nu_i}{\Omega_2}   {\rm ln } \hat\chi (\nu_1 - \nu_2 ; 
\Omega) {\rm ln } \hat\chi (\nu_1 - \nu_3 ;\Omega) \hat\chi (\nu_2 - \nu_3 ;\Omega) \non \\
= \frac{1}{(4 \pi)^3} \sum_{(m,n) \neq (0,0)} \frac{\Omega_2^3}{\vert m \Omega + n \vert^6}
= \frac{1}{(4 \pi)^3} E_3
(\Omega, \bar\Omega)^{SL(2,\mathbb{Z})} ,\eea
where we have used \C{expSL2}.

Thus, 
\be \label{calcmore}
\frac{(4\pi)^3}{4} \hat{I}_1 =  \int_{{\cal{F}}_L} \frac{d^2 \Omega}{\Omega_2^2} Z_{lat} E_3
(\Omega, \bar\Omega)^{SL(2,\mathbb{Z})} = \hat{I}_1^1 + \hat{I}_1^2 + \hat{I}_1^3 ,\ee
where $\hat{I}_1^1, \hat{I}_1^2$, and $\hat{I}_1^3$ are the contributions from the zero orbit, 
the non--degenerate orbits and the degenerate orbits of $SL(2,\mathbb{Z})$ respectively, as 
mentioned in the main text.

In order to evaluate \C{calcmore}, from \C{expSL2} we use the expression
\bea E_3 (\Omega,\bar{\Omega})^{SL(2,\mathbb{Z})} 
&=& 2 \zeta(6) \Omega_2^3 +  \frac{3\pi \zeta (5)}{4 \Omega_2^2} 
\non \\ &&+ \pi^3 \sqrt{\Omega_2} \sum_{m_1 \neq 0 ,m_2 \neq 0} 
\Big\vert \frac{m_1}{m_2} \Big\vert^{5/2} K_{5/2} (2\pi \Omega_2 \vert m_1 m_2
\vert) e^{2\pi i m_1 m_2 \Omega_1}. \eea
In doing the integrals, we 
frequently make use of the definition
\be K_s (x) = \frac{1}{2} \Big(  \frac{x}{2} \Big)^s \int_0^\infty \frac{dt}{t^{s+1}} e^{-t - x^2/4t} . \ee 

Integrating over the restricted fundamental domain 
${\cal{F}}_L$ of $SL(2,\mathbb{Z})$, we keep only the finite
terms in the limit $L \rightarrow \infty$.
The details of the calculation are very similar to~\cite{Basu:2007ru} and so we only mention the results.

{\bf{(i)}} The contribution from the zero orbit gives~\cite{Green:1999pv}
\be 
\hat{I}_1^1 = V_2 \int_{{\cal{F}}_L} \frac{d^2 \Omega}{\Omega_2^2} E_{3}
(\Omega, \bar\Omega)^{SL(2,\mathbb{Z})}  =0 , \ee
upto $L$ dependent terms. 

{\bf{(ii)}} The contribution from the non--degenerate orbits gives
\bea 
\hat{I}_1^2 &=& 2 V_2 \int_{-\infty}^\infty d \Omega_1 \int_0^\infty \frac{d\Omega_2}{\Omega_2^2} E_{3}
(\Omega, \bar\Omega)^{SL(2,\mathbb{Z})}
\sum_{k > j \geq 0, p \neq 0} e^{-2\pi i T kp - \frac{\pi T_2}{\Omega_2 U_2} 
\vert k\Omega + j + pU \vert^2} \non \\ &=& 2 \sqrt{T_2} E_{3}
(U, \bar{U})^{SL(2,\mathbb{Z})}\sum_{p \neq 0 ,k \neq 0} 
\Big\vert \frac{p}{k} \Big\vert^{5/2} K_{5/2} (2\pi T_2 \vert p k
\vert) e^{2\pi i p k T_1} ,
\eea
where we have also used
\be K_{1/2} (x) = \sqrt{\frac{\pi}{2 x}} e^{-x} , \quad 
K_{3/2} (x) = \sqrt{\frac{\pi}{2 x}} e^{-x} \Big( 1 + \frac{1}{x} \Big) , \quad 
K_{5/2} (x) = \sqrt{\frac{\pi}{2 x}} e^{-x} \Big( 1 + \frac{3}{x} + \frac{3}{x^2} \Big) ,\ee
and the identity
\be \frac{K_{1/2} (x+y)}{\sqrt{x+y}} + \frac{3\sqrt{x+y}}{xy} K_{3/2} (x+y) 
+ \frac{3(x+y)^{3/2}}{x^2 y^2} K_{5/2} (x+y) = {\sqrt{\frac{2xy}{\pi}}} \cdot \frac{K_{5/2} (x) 
K_{5/2} (y)}{x+y}.\ee

{\bf{(iii)}} The contribution from the degenerate orbits gives
\bea 
\hat{I}_1^3 &=& V_2 \int_{-1/2}^{1/2} d \Omega_1 \int_0^\infty \frac{d\Omega_2}{\Omega_2^2} E_{3}
(\Omega, \bar\Omega)^{SL(2,\mathbb{Z})} \sum_{(j,p) \neq (0,0)} 
e^{-\frac{\pi T_2}{\Omega_2 U_2} \vert j + pU \vert^2}\non \\
&=& \frac{2}{\pi^3} \Big( 2 \zeta (6) T_2^3
+ \frac{3 \pi \zeta (5) }{4 T_2^2} \Big)  E_{3} (U, \bar{U})^{SL(2,\mathbb{Z})} . \eea

Thus from \C{calcmore}, we get that
\be \hat{I}_1 = \frac{1}{8\pi^6} E_{3} (U, \bar{U})^{SL(2,\mathbb{Z})}
E_{3} (T, \bar{T})^{SL(2,\mathbb{Z})}. \ee

\subsection{Calculating $\hat{I}_2$}

We next evaluate \C{tortwo}, for which we use the representation
\be {\rm ln } \hat\chi (\nu ; \Omega) = \frac{\Omega_2}{4\pi} \sum_{n \neq 0} \frac{1}{n^2} 
e^{2\pi i n ({\rm Im}\nu)/\Omega_2} 
+ \frac{1}{4} \sum_{m \neq 0, k \in \mathbb{Z}} \frac{1}{\vert m \vert} e^{2\pi i m (k \Omega_1 + {\rm Re}\nu) 
-2\pi \Omega_2 \vert 
m \vert \vert k - ({\rm Im}\nu)/\Omega_2 \vert}\ee
for the scalar propagator on the torus. 
Again we write
\be \hat{I}_2 = \hat{I}_2^1 + \hat{I}_2^2 + \hat{I}_2^3 ,\ee
where $\hat{I}_2^1, \hat{I}_2^2$, and $\hat{I}_2^3$ are the contributions from the zero orbit, 
the non--degenerate orbits and the degenerate orbits of $SL(2,\mathbb{Z})$ respectively.

{\bf{(i)}} The contribution from the zero orbit gives~\cite{Green:1999pv}
\bea 
\hat{I}_2^1 &=& V_2 \int_{{\cal{F}}_L} \frac{d^2 \Omega}{\Omega_2^2}  \int_{\cal{T}} \prod_{i=1}^3 \frac{d^2 
\nu_i}{\Omega_2}  [ {\rm ln } \hat\chi (\nu_1 - \nu_2 ; 
\Omega)]^3 \non \\ &=& \frac{T_2}{32 \pi} \zeta (2) \zeta (3) ,\eea

upto $L$ dependent terms. This integral can be evaluated by using the Rankin--Selberg identity
to unfold the integration over the fundamental domain to the upper half plane, using the Poincare series representation
of the scalar propagator.

{\bf{(ii)}} The contribution from the non--degenerate orbits gives
\bea 
\hat{I}_2^2 &=& 2 V_2 \int_{-\infty}^\infty d \Omega_1 \int_0^\infty \frac{d\Omega_2}{\Omega_2^2} 
\int_{\cal{T}} \prod_{i=1}^3 \frac{d^2 
\nu_i}{\Omega_2}  [ {\rm ln } \hat\chi (\nu_1 - \nu_2 ; 
\Omega)]^3 \non \\ &&~~~~~~~~~~~~~~~~~~~~~~\times
\sum_{k > j \geq 0, p \neq 0} e^{-2\pi i T kp - \frac{\pi T_2}{\Omega_2 U_2} 
\vert k\Omega + j + pU \vert^2} \non \\ &=& \hat{I}_2^{2,1}
+ \hat{I}_2^{2,2} + \hat{I}_2^{2,3} ,\eea
where
\bea \hat{I}_2^{2,1} &=& 2 V_2 \sum_{m_1 \neq 0, m_2 \neq 0, m_3
\neq 0} \frac{\delta (m_1 +m_2+m_3)}{m_1^2 m_2^2 m_3^2}
\int_{-\infty}^\infty d \Omega_1 \int_0^\infty \frac{d\Omega_2}{\Omega_2^2} \Big( \frac{\Omega_2}{4\pi} 
\Big)^3 \non \\ &&~~~~~~~~~~~~~~~~~~~~~~\times
\sum_{k > j \geq 0, p \neq 0} e^{-2\pi i T kp - \frac{\pi T_2}{\Omega_2 U_2} 
\vert k\Omega + j + pU \vert^2} \non \\ &=& \frac{4 T_2 \zeta (6)}{(4\pi)^3} 
\int_{-\infty}^\infty d \Omega_1 \int_0^\infty d\Omega_2 \Omega_2 \sum_{k > j \geq 0, p \neq 0} 
e^{-2\pi i T kp - \frac{\pi T_2}{\Omega_2 U_2} \vert k\Omega + j + pU \vert^2} \non \\
&=& \frac{\zeta(6) U_2^3 \sqrt{T_2}}{16 \pi^3} \sum_{p \neq 0 ,k \neq 0} 
\Big\vert \frac{p}{k} \Big\vert^{5/2} K_{5/2} (2\pi T_2 \vert p k
\vert) e^{2\pi i p k T_1},\eea

\bea  \label{diffcal2}
\hat{I}_2^{2,2} &=& \frac{3 V_2}{32\pi} 
\int_{-\infty}^\infty d \Omega_1 \int_0^\infty \frac{d\Omega_2}{\Omega_2} 
\int_0^1 dx \int_0^1 dy \sum_{m \neq 0} \frac{1}{m^2} e^{2\pi i m (x-y)}\non \\ &&\times
\sum_{n \neq 0} \frac{1}{n^2} \sum_{(r_1,r_2) \in \mathbb{Z}} 
e^{2\pi i n (r_1 - r_2) \Omega_1 - 2\pi \vert n \vert \Omega_2 (\vert r_1 - (x-y)\vert 
+ \vert r_2 - (x-y)\vert)} \non \\ && \times
\sum_{k > j \geq 0, p \neq 0} e^{-2\pi i T kp - \frac{\pi T_2}{\Omega_2 U_2} 
\vert k\Omega + j + pU \vert^2} \non \eea
\bea &&= \frac{3 V_2}{32\pi^2} 
\int_{-\infty}^\infty d \Omega_1 \int_0^\infty d\Omega_2 \sum_{n \neq 0} \frac{1}{n^2} 
\non \\ &&\times \sum_{m \neq 0} \frac{1}{\vert m \vert^3} \frac{d}{d\Omega_2} \Big\{ {\rm tan}^{-1}
\Big( \frac{2\Omega_2 \vert n \vert}{\vert m \vert} \Big)\Big\} 
\frac{(1-e^{-4\pi \vert n \vert \Omega_2})}{(1 -2 {\rm cos} (2\pi n \Omega_1)e^{-2\pi \vert n \vert \Omega_2}
+e^{-4\pi \vert n \vert \Omega_2})} \non \\ && \times
\sum_{k > j \geq 0, p \neq 0} e^{-2\pi i T kp - \frac{\pi T_2}{\Omega_2 U_2} 
\vert k\Omega + j + pU \vert^2} .\eea

Now using the representation~\cite{GradRyz} 
\bea \label{exptan} {\rm tan}^{-1} x &=& \frac{\pi}{2} - \sum_{k=0}^\infty \frac{(-1)^k}{(2k+1) x^{2k+1}} \non \\
&=& \frac{\pi}{2} - \frac{1}{x} + \frac{1}{3x^3} + \ldots,\eea
we see that only the $k=0$ and $k=1$ terms in \C{exptan} contribute to \C{diffcal2} while doing the sum over $m$. 
While the constant term trivially vanishes, the terms for $k \geq 2$ vanish on doing the sum over $m$, 
because $\zeta (-2p) =0$ for all positive integers $p$. Thus, using $\zeta (0) = -1/2$, we get that

\bea \label{verylongexp}
\hat{I}_2^{2,2} &=& \frac{3 V_2}{32\pi^2} \int_{-\infty}^\infty d \Omega_1 \int_0^\infty d\Omega_2  
\sum_{n \neq 0} \frac{1}{\vert n \vert^3} \Big( \frac{\zeta (2)}{\Omega_2^2} + \frac{1}{8 n^2 \Omega_2^4}\Big) 
\non \\ && \times \Big\{ 1 + \frac{2 e^{-2\pi \vert n \vert \Omega_2} ({\rm cos} (2\pi n \Omega_1)
- e^{-2\pi \vert n \vert \Omega_2})}{1 -2 {\rm cos} (2\pi n \Omega_1)e^{-2\pi \vert n \vert \Omega_2}
+e^{-4\pi \vert n \vert \Omega_2}} \Big\}
\non \\ && \times \sum_{k > j \geq 0, p \neq 0} e^{-2\pi i T kp - \frac{\pi T_2}{\Omega_2 U_2} 
\vert k\Omega + j + pU \vert^2}. \eea

Now in \C{verylongexp}, the terms in $\{ \ldots \}$ are 1, and another term which exponentially decreases as
$\Omega_2 \rightarrow \infty$. We call these two contributions $\hat{I}_2^{2,2} (1{\rm  only})$
and $\hat{I}_2^{2,2} ({\rm not} 1)$ respectively. 

The term involving 1 gives us 
\bea \hat{I}_2^{2,2} (1{\rm  only})&=& \frac{3\sqrt{T_2}}{32\pi^2} \Big[ 2 \zeta (2) \zeta (3) 
\sum_{p \neq 0 ,k \neq 0} \Big\vert \frac{p}{k} \Big\vert^{1/2} K_{1/2} (2\pi T_2 \vert p k
\vert) e^{2\pi i p k T_1} \non \\ &&+ \frac{\zeta (5)}{4 U_2^2} \sum_{p \neq 0 ,k \neq 0} 
\Big\vert \frac{p}{k} \Big\vert^{5/2} K_{5/2} (2\pi T_2 \vert p k
\vert) e^{2\pi i p k T_1} \Big], \eea
while $\hat{I}_2^{2,2} ({\rm not} 1)$ can be expanded in a power series in $e^{-2\pi \vert n \vert \Omega_2}$ for large
$\Omega_2$, and integrated term by term. This gives a far more complicated expression which we shall return to later.

Finally, the remaining expression is given by
\bea \label{needmorexp}
\hat{I}_2^{2,3} &=& \frac{V_2}{32} \int_{-\infty}^\infty d \Omega_1 \int_0^\infty \frac{d\Omega_2}{\Omega_2^2}
\int_0^1 dx \int_0^1 dy \non \\ && \times \sum_{m_1,m_2,m_2 \neq 0 ; l_1, l_2, l_3 \in \mathbb{Z}}
\frac{\delta (\sum_i m_i)}{\vert m_1 m_2 m_3 \vert}
e^{2\pi i m_i l_i \Omega_1 - 2\pi \Omega_2 \vert m_i \vert 
\vert l_i - (x-y) \vert}\non \\ &&\times \sum_{k > j \geq 0, p \neq 0} e^{-2\pi i T kp - \frac{\pi T_2}{\Omega_2 U_2} 
\vert k\Omega + j + pU \vert^2}.\eea
We shall also return to this expression later.

{\bf{(iii)}} The contribution from the degenerate orbits gives 
\bea \hat{I}_2^3 &=& V_2 \int_{-1/2}^{1/2} d \Omega_1 \int_0^L \frac{d\Omega_2}{\Omega_2^2} 
 \int_{\cal{T}} \prod_{i=1}^3 \frac{d^2 
\nu_i}{\Omega_2}  [ {\rm ln } \hat\chi (\nu_1 - \nu_2 ; 
\Omega)]^3 
\sum_{(j,p) \neq (0,0)} 
e^{-\frac{\pi T_2}{\Omega_2 U_2} \vert j + pU \vert^2} \non \\ &=& \hat{I}_2^{3,1}
+ \hat{I}_2^{3,2} + \hat{I}_2^{3,3} ,\eea

where
\bea \hat{I}_2^{3,1} &=& V_2 \int_0^L \frac{d\Omega_2}{\Omega_2^2} \int _0^1 dx \int_0^1
dy \Big( \frac{\Omega_2}{4\pi} \Big)^3 \Big( \sum_{m \neq 0} \frac{1}{m^2} e^{2\pi i m (x-y)} \Big)^3
\sum_{(j,p) \neq (0,0)} 
e^{-\frac{\pi T_2}{\Omega_2 U_2} \vert j + pU \vert^2} \non \\ &=& \frac{T_2^3 E_{3} (U, 
\bar{U})^{SL(2,\mathbb{Z})}}{32\pi^6} \sum_{m\neq 0, n\neq 0, p\neq 0} \frac{\delta (m+n+p)}{m^2 n^2 p^2} \non \\
&=& \frac{\zeta (6)}{16 \pi^6} T_2^3 E_{3} (U, \bar{U})^{SL(2,\mathbb{Z})}.\eea

Also
\bea \label{diffcal}
\hat{I}_2^{3,2} = \frac{3 V_2}{64 \pi} \int_0^L \frac{d\Omega_2}{\Omega_2} \int _0^1 dx \int_0^1 dy
\sum_{m \neq 0, n \neq 0, k \in \mathbb{Z}} \frac{1}{m^2 n^2} e^{2\pi i m (x-y) -4\pi \Omega_2
\vert n \vert \vert k - (x-y) \vert} \non \\ \times \sum_{(j,p) \neq (0,0)} 
e^{-\frac{\pi T_2}{\Omega_2 U_2} \vert j + pU \vert^2} \non \eea
\bea = \frac{3 V_2}{64 \pi^2} \sum_{m \neq 0, n \neq 0} \frac{1}{\vert m \vert^3 n^2} \int_0^\infty dx \frac{d}{dx}
\Big\{ {\rm tan}^{-1} \Big( \frac{2 \vert n \vert x}{\vert m \vert}\Big) \Big\} \sum_{(j,p) \neq (0,0)} 
e^{-\frac{\pi T_2}{x U_2} \vert j + pU \vert^2}. \eea

Using the representation \C{exptan}, once again we see that only the $k=0$ and $k=1$ terms in \C{exptan} contribute to 
\C{diffcal}. Thus we get that
\be \hat{I}_2^{3,2} = \frac{3}{32 \pi^3} \zeta (2) \zeta (3) E_1 (U, \bar{U})^{SL(2,\mathbb{Z})}
+ \frac{3\zeta (5)}{128 \pi^5 T_2^2} E_{3} (U, \bar{U})^{SL(2,\mathbb{Z})}.\ee

Also we have that
\bea \label{reallong}
\hat{I}_2^{3,3} &=& \frac{V_2}{64} \int_0^L \frac{d\Omega_2}{\Omega_2^2}
\int_0^1 dx \int_0^1 dy \non \\ && \times \sum_{m_1,m_2,m_2 \neq 0 ; l_1, l_2, l_3 \in \mathbb{Z}}
\frac{\delta (\sum_i m_i) \delta (\sum_i l_i m_i)}{\vert m_1 m_2 m_3 \vert}
e^{- 2\pi \Omega_2 \vert m_i \vert 
\vert l_i - (x-y) \vert}\non \\ &&\times \sum_{(j,p) \neq (0,0)} 
e^{-\frac{\pi T_2}{\Omega_2 U_2} \vert j + pU \vert^2}.\eea
Although \C{reallong} is a complicated expression, it is not difficult to see that the integrand goes as $O(e^{-\Omega_2})$
as $\Omega_2 \rightarrow \infty$, and does not involve any power law suppressed terms. 
Thus we have that
\be \hat{I}_2^{3,3} = T_2 \sum_{M,N} g_{MN} \int_0^\infty d\Omega_2 \Omega_2^{-M} e^{-N\Omega_2} \sum_{(j,p) \neq (0,0)} 
e^{-\frac{\pi T_2}{\Omega_2 U_2} \vert j + pU \vert^2},\ee
where $g_{M,N}$ are unspecified functions of $M$ and $N$, $M$ is an integer, and $N$ is non--zero.

Let us denote the terms independent of $T_1$ and $U_1$
in the various expressions as perturbative in $T$ and $U$ respectively (not to be confused with string perturbation 
theory). Thus $\hat{I}_2^{3,3}$ is perturbative in $T$, but has a non--trivial dependence on $U_1$. First
let us consider the terms in $\hat{I}_2^{3,3}$ which are perturbative in $U$ as well. In order to
do this, we use the relation
\bea \label{relpoisson}
&&\sum_{(j,p) \neq (0,0)} e^{-\frac{\pi T_2}{\Omega_2 U_2} \vert j + pU \vert^2} \non \\
&&= \sum_{j\neq 0} e^{-\frac{\pi T_2 \vert j \vert^2}{U_2 \Omega_2}} + \sqrt{\frac{U_2 \Omega_2}{T_2}}
\sum_{p\neq 0} e^{-\frac{\pi p^2 T_2 U_2}{\Omega_2}}  
 + \sqrt{\frac{U_2 \Omega_2}{T_2}} \sum_{p\neq 0, \hat{j} \neq 0} e^{2\pi i p \hat{j} \bar{U}
-\frac{\pi U_2}{T_2 \Omega_2} (pT_2 + \hat{j} \Omega_2)^2}.\eea
We now outline the principal steps to deduce the various terms on the right hand side of \C{relpoisson}.
The first term is obtained by setting $p=0$, while to obtain the remaining terms which have $p \neq 0$, 
we Poisson resum on $j$ to go to the variable $\hat{j}$. The second term is given by the $\hat{j} =0$
contribution, while the third term has $\hat{j} \neq 0$. 
Thus the first two terms in \C{relpoisson} give the perturbative contributions.                

This gives
\bea \label{finpert}
&&\hat{I}_2^{3,3} ({\rm pert}) \non \\ &&= 2 \sum_{M,N} \frac{g_{MN}}{(\pi N^{-1})^{(M-1)/2} T_2^{(M-3)/2}}
\sum_{j \neq 0} \Big( \frac{U_2}{\vert j \vert^2} \Big)^{(M-1)/2} K_{M-1} (2 \vert j \vert
\sqrt{\frac{\pi N T_2}{U_2}}) \non \\ && + 2 \sum_{M,N} \frac{g_{MN}}{(\pi N^{-1})^{(2M -3)/4}
(U_2 T_2)^{(2M -5)/4}} \sum_{p\neq 0} \frac{1}{\vert p \vert^{M - 3/2}} K_{M- 3/2} (2\vert
p \vert \sqrt{\pi N T_2 U_2}). \eea

We now fix $\hat{I}_2^{3,3} ({\rm pert})$ using the constraint that the amplitude must be the same in the two
type II string theories. Note that the perturbative parts come only from the zero orbit and the degenerate
orbit contributions to the amplitude. Thus from the perturbative contributions already calculated
in $\hat{I}_2^1$, $\hat{I}_2^{3,1}$, and $\hat{I}_2^{3,2}$ we see that $\hat{I}_2^{3,3} ({\rm pert})$
must contain
\be \label{restpert}
-\frac{3\zeta (2) \zeta (3)}{32 \pi^2} {\rm ln} T_2 .\ee

We now argue that there are no other perturbative contributions to $\hat{I}_2^{3,3} ({\rm pert})$.
Suppose there are other such contributions apart from \C{restpert}: because these are the only 
remaining ones, and they must be symmetric under interchange of $U_2$ and $T_2$, they must be of the form
\be \label{restpert2} h (U_2) + h (T_2) + \sum_i r_i (U_2) r_i (T_2). \ee

Thus the derivative with respect to $U_2$ of the total perturbative contributions \C{restpert} 
and \C{restpert2} is given by
\be  \label{restpert3} \frac{\p h (U_2)}{\p U_2} + \sum_i \frac{\p r_i (U_2)}{\p U_2} r_i (T_2). \ee
Consider the large $U_2$ limit of \C{restpert3}. Let $h (U_2) \sim U_2^\lambda$, and $r_i (U_2) \sim
U_2^{\lambda_i}$ for large $U_2$\footnote{Assuming a more general behavior of the form 
$h (U_2) \sim U_2^\lambda ({\rm ln} U_2)^{\hat\lambda}$, $r_i (U_2) \sim U_2^{\lambda_i}
({\rm ln} U_2)^{\hat\lambda_i}$ does not change the conclusions below.}. Thus \C{restpert3} has to contain a term
\be \label{restpert4} \lambda U_2^{\lambda -1} + T_2 \sum_i \lambda_i^2 (U_2 T_2)^{\lambda_i - 1} \ee
at large $U_2$. Now consider the large $U_2$ behavior of the $U_2$ 
derivative of \C{finpert}. For large $x$,
using the relation
\be K_s (x) \sim \sqrt{ \frac{\pi}{2x}} e^{-x},\ee
we see that the second term does not contribute. On the other hand, 
for small $x$ using the relations
\be K_0 (x) \sim - {\rm ln} ~x ; \quad K_m (x) \sim \frac{\Gamma (m)}{2}  
\Big( \frac{x}{2} \Big)^{-m}, m > 0, \ee
from the first term, we get that
\be \frac{\p\hat{I}_2^{3,3}}{\p U_2} ({\rm pert}) \sim  -\frac{T_2}{U_2}
\sum_{N} g_{1N} + 2 \sum_{M > 1,N} 
\frac{g_{MN} \Gamma (M) \zeta (2M -2)}{\pi^{M-1}} \Big( \frac{U_2}{T_2} 
\Big)^{M-2}, \ee
which can never be of the form \C{restpert4}. Thus 
\be \hat{I}_2^{3,3} ({\rm pert}) = -\frac{3\zeta (2) \zeta (3)}{32 \pi^2} {\rm ln} T_2 . \ee
This contribution has a logarithmic dependence on $T_2$ and must arise from the infinite sum over $N$
in \C{finpert}. Any constant
term in $\hat{I}_2^{3,3} ({\rm pert})$ can be absorbed in the regularization of the infrared divergences. 
Thus we see that
\bea \label{finalpert}
\hat{I}_2^{\rm pert} &=& \frac{1}{32 \pi^6} E_{3} (U, \bar{U})^{SL(2,\mathbb{Z})}_{\rm pert}
E_{3} (T, \bar{T})^{SL(2,\mathbb{Z})}_{\rm pert} \non \\ &&+ \frac{3}{32\pi^3} \zeta (2) \zeta (3) \Big( 
E_{1} (U, \bar{U})^{SL(2,\mathbb{Z})}_{\rm pert} + E_{1} (T, \bar{T})^{SL(2,\mathbb{Z})}_{\rm pert}
\Big).\eea
Extending \C{finalpert} to its non--perturbative completion, we get that
\bea \label{finalnonpert}
\hat{I}_2 &=& \frac{1}{32 \pi^6} E_{3} (U, \bar{U})^{SL(2,\mathbb{Z})}
E_{3} (T, \bar{T})^{SL(2,\mathbb{Z})} \non \\ &&+ \frac{3}{32\pi^3} \zeta (2) \zeta (3) \Big( 
E_{1} (U, \bar{U})^{SL(2,\mathbb{Z})} + E_{1} (T, \bar{T})^{SL(2,\mathbb{Z})}
\Big).\eea

Considering the various non--perturbative contributions that have already been calculated in $\hat{I}_2^{2,1}$,
$\hat{I}_2^{2,2} ({\rm 1 only})$, $\hat{I}_2^{3,1}$ and $\hat{I}_2^{3,2}$, we get all the terms in 
\C{finalnonpert} with the precise coefficients apart from just one term. This term is
\be \label{lastnonpert}
\frac{\sqrt{U_2 T_2}}{32} \sum_{p \neq 0 ,k \neq 0} 
\Big\vert \frac{p}{k} \Big\vert^{5/2} K_{5/2} (2\pi T_2 \vert p k
\vert) e^{2\pi i p k T_1} \sum_{m \neq 0 ,n \neq 0} 
\Big\vert \frac{m}{n} \Big\vert^{5/2} K_{5/2} (2\pi T_2 \vert m n
\vert) e^{2\pi i m n U_1}. \ee
Now \C{lastnonpert} depends on $T_1$, and thus cannot be obtained from $\hat{I}_2^{3,3}$. Thus
\be \hat{I}_2^{3,3} ({\rm non-pert}) = 0 .\ee 
So \C{lastnonpert} must be obtained from the only remaining contributions leading to  
\bea && \hat{I}_2^{2,2} ({\rm not1}) +\hat{I}_2^{2,3}  \\ 
&&= \frac{\sqrt{U_2 T_2}}{32} \sum_{p \neq 0 ,k \neq 0} 
\Big\vert \frac{p}{k} \Big\vert^{5/2} K_{5/2} (2\pi T_2 \vert p k
\vert) e^{2\pi i p k T_1} \sum_{m \neq 0 ,n \neq 0} 
\Big\vert \frac{m}{n} \Big\vert^{5/2} K_{5/2} (2\pi T_2 \vert m n
\vert) e^{2\pi i m n U_1}.\non \eea

This concludes the calculation of the torus amplitude. We have obtained some parts of the amplitude
based on consistency and heuristic arguments, but have not explicitly calculated 
those contributions. It would be nice to calculate them explicitly. In the next appendix,
we provide some more evidence that the extra contributions in \C{restpert2} vanish.

\section{A self--consistency check for the torus amplitude} 

In the previous section, we have calculated the four graviton amplitude on the torus. Some parts of the
amplitude were obtained using indirect arguments and not by explicit calculations. We now show that
the answer we got is consistent with the structure we have proposed for the modular form.

We mentioned that there can be additional contributions to the torus amplitude given by \C{restpert2}. Let
$h (T, \bar{T})_{\rm pert} \equiv h (T_2)$. We now show that $h (T , \bar{T}) =0$ based on very different 
considerations compared to the previous discussion. This contribution yields an additional term
$\mu h (T , \bar{T})$ to \C{pertexact}.
Repeating the arguments as before, we get back the results of section 4, alongwith an extra
equation given by
\be \Delta_{{SL(2,\mathbb{Z})}_T} h (T , \bar{T}) = 12 h (T , \bar{T}).\ee
This is, of course, solved by $E_4 (T,\bar{T})^{SL(2,\mathbb{Z})}$. Thus adding 
$E_4 (T,\bar{T})^{SL(2,\mathbb{Z})} +E_4 (U,\bar{U})^{SL(2,\mathbb{Z})}$ to the torus amplitude, we see that
in nine dimensions this leads to a divergent term 
\be \label{wrongdep} 2 \zeta (8) l_s^5 \int d^9 x \sqrt{-g_9} 
\Big( r_B^4 + \frac{1}{r_B^4} \Big)  r_\infty^3  D^6 {\mathcal{R}}^4 .\ee
However, from \C{divnine} we see that the divergence needed to produce threshold singularities 
should go as $r_\infty^5$, and is also independent of $r_B$. This has a different behavior than \C{wrongdep},
thus $h (T , \bar{T})=0$.

Also the $T_2^4$ dependence of the torus amplitude at large $T_2$ is inconsistent with the large $T_2$ scaling 
behavior of the genus two and three amplitudes based on considerations of degeneration limits of the
Riemann surfaces as discussed before, which leads to the same conclusion. 

The remaining terms in \C{restpert2} give an additional perturbative (in the string coupling) 
contribution to the proposed modular form 
\be \mu \sum_i r_i (T,\bar{T})  r_i (U,\bar{U}) ,\ee
where $ r_i (T, \bar{T})_{\rm pert} \equiv r_i (T_2)$. Thus $r_i (U,\bar{U})$ must be an $SL(2,\mathbb{Z})_U$
invariant modular form, while $\mu r_i (T,\bar{T})$ must get enhanced to an $SL(3,\mathbb{Z})_M$
invariant modular form $r_i (M)$. Now using the symmetry under interchange of $U$ and $T$, we conclude
that $r_i (M)$ receives only one perturbative contribution at genus zero, and instanton corrections.

On the other hand, we know that $r_i (M)$ must satisfy the Laplace equation, or a Poisson equation 
on moduli space. If it satisfies the Laplace equation, it will have two perturbative contributions,
contradicting the statement above. If it satisfies a Poisson equation,
considerations of supersymmetry constrain the source term to involve the modular form for 
the ${\mathcal{R}}^4$ interaction, namely, $E_{3/2} (M)^{SL(3,\mathbb{Z})}$, which has a genus zero 
and a genus one contribution. Thus the solution of the Poisson equation will have more than one 
perturbative contribution, again contradicting the statement above. Thus $r_i (M)=0$, and
\C{restpert2} vanishes.   






\providecommand{\href}[2]{#2}\begingroup\raggedright\endgroup

\end{document}